\documentclass[twocolumn,amsmath,amssymb,aps,showpacs,graphicx]{revtex4-2}

\usepackage{graphicx}
\usepackage{dcolumn}
\usepackage{bm}
\usepackage{lineno, hyperref}
\usepackage{natbib}
\usepackage{multirow}
\begin{document}

\title{Unusual spin-triplet superconductivity in monolayer graphene}
\author{V. Apinyan} 
\altaffiliation[e-mail:]{v.apinyan@intibs.pl}
\author{M. Sahakyan} 
\affiliation{Institute of Low Temperature and Structure Research, Polish Academy of Sciences\\ 50-422, Wroc\l{}aw 2, Poland 
}
%
\begin{abstract}
In this paper we consider the phonons in monolayer graphene and we show the possibility for the spin-triplet superconducting excitations states by discretizing the single-particle excitations near Fermi wave vector. The molonayer graphene was supposed to be exposed under the influence of the external gate-potential and the local Coulomb interaction effects have been taken into account at each lattice site position in the monolayer. A sufficiently large temperature domain was found, where the superconducting order parameter is not vanishing. Corresponding to this, at the surprisingly high temperature limit, we obtain a narrow domain of the electron-phonon coupling parameter $\lambda_{\rm eff}$, emphasizing the superconducting state. We discuss the localizing role of Hubbard-$U$ interaction and the effects external gate potential on the calculated physical parameters in the system. We explain the importance of the chemical potential in the formation of the superconducting state. We show the existence of a large superconducting band-gap in the system even in the case of the absence of the applied electric field potential.     
\end{abstract}     



\maketitle
%
\section{\label{sec:Section_1} Introduction}
%
The monolayer graphene has been the subject of many interesting research investigations since its discovery \cite{cite_1,cite_2}. The band-gap opening in graphene band structure have been acheaived by using different techniques such as hydrogenation \cite{cite_3}, substrate effect \cite{cite_4}, layered structures and locally induced electric field via self-assembling molecular network \cite{cite_5}. Among the interesting electronic conduction and band structure properties, the monolayer grahene shows a two-dimensional (2D) superconductivity (SC) when intercalating with Ca \cite{cite_6}, and the superconducting state is then due to the low-energy vibrations of the intercalated atoms \cite{cite_7}. The enhancement of superconductivity in graphene has been shown in Ref.\cite{cite_8}, where the pairing instability occurs after feeding the experimental Fermi surface topology into the calculations of the electronic band structure and taking into account the influence of the extended van Hove singularities vHS \cite{cite_8,cite_9}.  It was supposed that the electron-phonon interaction adds an additional mass term near the vHS and contribute to singularities in the density of states. Therefore, the enhancing SC properties, related to the behavior of vHS, are the results of competition between electron-phonon and electron-electron interactions in graphene \cite{cite_10}. Recently, it has been shown \cite{cite_11}, via the Angle-Resolved Photoemission Spectroscopy measurements, that monolayer graphene becomes superconducting after depositing the Li adatoms at the very low temperatures of order $5.9$ K. This was theoretically proposed many years ago concerning the alkali adatom deposition \cite{cite_12}. Various mechanisms have been proposed theoretically, for understanding the origin of SC in monolayer graphene \cite{cite_13, cite_14, cite_15, cite_16, cite_17, cite_18}. It is argued, in Ref.\cite{cite_14}, that the realization of the SC in pure monolayer graphene is a cumbersome task because of the vanishing density of states at Fermi level and the SC state appears only when shifting Fermi level by doping graphene with metallic atoms. Moreover, it was shown in Ref.\onlinecite{cite_15} that for the undoped case, there exists a quantum critical point which leads to the suppression of the superconducting order below some concrete value of the pairing interaction strength. 

Various interesting investigations have been dedicated for study the electron-phonon coupling in graphene \cite{cite_19, cite_20, cite_21, cite_22, cite_23, cite_24}.
The results, obtained from Renishaw Raman spectroscopy measurements\cite{cite_19}, show that the main contribution into the Raman spectrum comes from the scalar two-phonon peaks from the vicinity of Dirac $K$-point in the Brillouin zone. The electron-phonon coupling constant found from the experimental Raman techniques is nearly $2.5$ times larger than the values calculated from the \textit{ab initio} density functional theory \cite{cite_25}. On the other hand, Coulomb interaction between the electrons in graphene leads to a strong enhancement of the electron-phonon coupling in the neighborhood of $K$-point \cite{cite_17}. Recently \cite{cite_20}, the Resonance Raman spectroscopy techniques has been elaborated to estimate the electron-phonon coupling energy in graphene at the close vicinity of Dirac $K$-point, exploring the resonant regions of the electron and phonon dispersions, ranging from $3.06$ to $1.16$ eV on the quasi-particle excitation energy scale. The authors have found a huge enhancement of the intensity peaks at the $K$-point in good agreement with the results in \cite{cite_19}. Meanwhile, the interpretation of the intensity peak is done in terms of the proximity of the ${\bf{q}}$-excitations and the $K$-point. Moreover, the 2D line-width of the excitation peak-region in \cite{cite_19} was addressed after comparing the results with the \textit{ab initio} calculations.       

It appears that another interesting mechanism for creating the SC state in graphene is the high temperature annealing of graphene \cite{cite_26, cite_27, cite_28, cite_29} which favors the creation of external adatoms and consequently the possible superconducting pairing. This mechanism is especially important for creating an unintentional doping effects in the monolayer graphene.  

In the present paper, we consider a single layer suspended graphene, exposed to the external electric field gate-potential and we examine the electron-phonon coupling in terms of two-particle ${\bf{q}}$-excitations at the vicinity of Fermi level. The contact potentials have been applied to different sublattice sites in graphene. We obtain the superconducting order parameter after detailed field-integration of phonons and we derive the expression of the effective electron-electron interaction in graphene. The chemical potential, which plays the essential role in the formation of new quasiparticles has been calculated exactly via the equation for the on-site average occupation (half-filling or partial-filling) number of particles. We show that the charge neutrality occurs only in the case of the unbiased graphene layer. At the very high temperature limit, a narrow region of spin-triplet superconducting phase has been obtained as a function of the electron-phonon coupling constant. We show that the appearance of SC is a consequence of discretization of single-particle ${\bf{q}}$-excitations at Fermi level. We obtain the temperature dependence of the superconducting order parameter. Moreover, the role of the local Hubbard-$U$ interaction and external gate-potential has been discussed and the electronic band-structure has been calculated.

The paper is organized as follow: In Section \ref{sec:Section_2} we introduce Hamiltonian of the system and we discuss the electron-phonon coupling mechanism. In Section \ref{sec:Section_3}, we give, in details, the process of integration of the phonon-field and we obtain the effective electron-electron action. The numerical results have been embedded in the Section \ref{sec:Section_4}, within the discussion of obtained results. In the Section \ref{sec:Section_5}, we give a short conclusion to our paper. The Appendix in Section \ref{sec:Section_6} is devoted to the calculation of some important coefficients entering in the system of self-consistent equations for the superconducting gap.
%
\section{\label{sec:Section_2} Hamiltonian of the graphene}
\begin{figure}
	\includegraphics[scale=0.4]{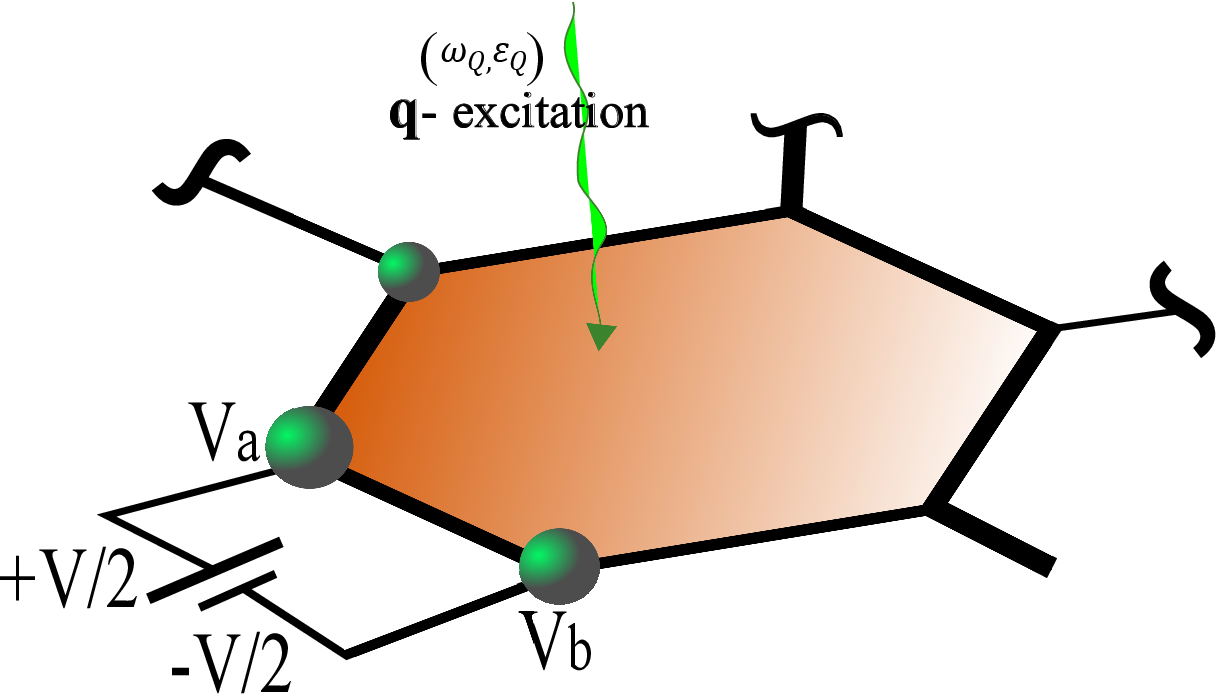}
	\caption{\label{fig:Fig_1}(Color online) The monolayer graphene exposed to the influence of the external electric field potential $V$. The green arrow shows the external excitation wave with the given wave vector ${\bf{Q}}$ and energy $\varepsilon_{{\bf{Q}}}=\hbar\Omega_{Q}$. We suppose that the voltage drop at the sublattice site $a$ is $V_{a}=+V/2$ and at the sublattice site $b$ is $V_{b}=-V/2$.}
\end{figure}

%
We describe the system of single layer, non-interacting, graphene via tight-binding model
\begin{eqnarray}
{\cal{\hat{H}}}_{\rm t-B}&=&-\gamma_{0}\sum_{\left\langle {\bf{r}}{\bf{r}}'\right\rangle, \sigma}\left(\hat{a}^{\dag}_{\sigma}\left({\bf{r}}\right)\hat{b}_{\sigma}\left({\bf{r}}'\right)+h.c.\right)
\nonumber\\
&&-\mu\sum_{{\bf{r}}}\left(\hat{n}_{a}\left({\bf{r}}\right)+\hat{n}_{b}\left({\bf{r}}\right)\right),
	\label{Equation_1}
\end{eqnarray}
where $\gamma_0$ is the hopping amplitude between the nearest neighbor lattice sites ${\bf{r}}$ and ${\bf{r}}'$. The value of this parameter is $\gamma_0\sim 3$ eV, which was found experimentally in Refs.\onlinecite{cite_30, cite_31}. The operators $\hat{a}^{\dag}_{\sigma}\left({\bf{r}}\right)$ ($\hat{a}_{\sigma}\left({\bf{r}}\right)$) and $\hat{b}^{\dag}_{\sigma}\left({\bf{r}}\right)$ ($\hat{b}_{\sigma}\left({\bf{r}}\right)$) are the electron creation (annihilation) operators (with spin $\sigma$) at the given lattice site ${\bf{r}}$. The last term indicates the link between the chemical potential and the single-particle densities $\hat{n}_{a}\left({\bf{r}}\right)$ and $\hat{n}_{b}\left({\bf{r}}\right)$, which are defined as 
\begin{eqnarray}
\hat{n}_{\eta}\left({\bf{r}}\right)=\sum_{{\bf{r}},\sigma}\hat{\eta}^{\dag}_{\sigma}\left({\bf{r}}\right)\hat{\eta}_{\sigma}\left({\bf{r}}\right)
	\label{Equation_2}
\end{eqnarray}
with $\hat{\eta}=\hat{a},\hat{b}$.
Here, we add to Hamiltonian in Eq.(\ref{Equation_1}) also terms describing the local on-site interaction $U$ and the external gate-potential $V$
\begin{eqnarray}
&&{\cal{\hat{H}}}_{\rm U-V}=U\sum_{{\bf{r}},\eta=a,b}\hat{n}_{\eta\uparrow}\left({\bf{r}}\right)\hat{n}_{\eta\downarrow}\left({\bf{r}}\right)
+\sum_{{\bf{r}},\eta=a,b}V_{\eta}\hat{n}_{\eta}\left({\bf{r}}\right),
\nonumber\\
\label{Equation_3}
\end{eqnarray}
where the first term is the subject of the usual Hubbard interaction, and the second term describes the coupling between the external electric field potential and the single-particle densities at the sites $a$ and $b$ of the graphene lattice. We put here $V_{a}=-V_{b}=V/2$, where $V$ is the applied electric field potential (see in see in Fig.~\ref{fig:Fig_1}). 
%
\subsection{\label{sec:Section_2_1} The electron-phonon interaction}
%
Due to the small displacement of atoms with a vector ${\bf{R}}$, the electron concentration imbalance (described via the operator $\delta{\hat{n}}\left({\bf{r}},{\bf{r}}'\right)$) was created in the system, which leads to the modification of the total energy of electrons in the system by a quantity $\delta{\hat{E}}$ 
\begin{eqnarray}  
\delta{\hat{E}}=\int\int{d{\bf{r}}d{\bf{r}}'}\delta{\hat{n}}\left({\bf{r}},{\bf{r}}'\right)W_{\rm el-ion}\left({\bf{r}}-{\bf{r}}'\right),
\label{Equation_4}
\end{eqnarray}
where $W_{\rm el-ion}$ is the electron-ion interaction potential when we have the static ionic background with the charge $Q_{\rm ion}$. We have $W_{\rm el-ion}=Q_{\rm el}Q_{\rm ion}/|{\bf{r}}-{\bf{r}}'|$, and the corresponding electronic ($Q_{\rm el}$) and ionic ($Q_{\rm ion}$) charges are defined as $Q_{\rm e}\left({\bf{r}}\right)=-en\left({\bf{r}}\right)=-e|\psi_{\rm el}\left({\bf{r}}\right)|^{2}$ and $Q_{ion}=-Zen_{ion}=-Ze|\psi_{\rm atom}\left({\bf{r}}\right)|^{2}$ (with $\psi_{\rm el}\left({\bf{r}}\right)$ and $\psi_{\rm atom}\left({\bf{r}}\right)$ being the electronic and ionic wave functions). The charge density imbalance $\delta{\hat{n}}\left({\bf{r}},{\bf{r}}'\right)$ could be expressed in terms of small atomic displacement vector ${\bf{R}}$: $\delta{\hat{n}}\left({\bf{r}},{\bf{r}}'\right)=\hat{n}\left({\bf{r}}\right)-\hat{n}\left({\bf{r}}'\right)=-{\bf{R}}\partial{\hat{n}}\left({\bf{r}}'\right)/\partial{\bf{r}}'=-{\rm div}{\hat{{\bf{R}}}}\left({\bf{r}}'\right)$, where $\hat{{\bf{R}}}\left({\bf{r}}'\right)={\bf{R}}{n}\left({\bf{r}}'\right)$ is the atomic displacement operator corresponding the lattice site position ${\bf{r}}'$. For the total electron energy modification, we get
\begin{eqnarray}  
\delta{\hat{E}}=-ea^{2}\int{d{\bf{r}}{n}\left({\bf{r}}\right){\rm div}{\hat{{\bf{P}}}}\left({\bf{r}}\right)}.
\label{Equation_5}
\end{eqnarray}
Here, ${\hat{{\bf{P}}}}\left({\bf{r}}\right)$ is the atomic polarization operator: ${\hat{{\bf{P}}}}\left({\bf{r}}\right)=Zen_{ion}{\hat{{\bf{R}}}}\left({\bf{r}}\right)$, and $a$ is the constant of order of atomic sites separation $a_{0}$ \cite{cite_32}. Furthermore, we define the atomic displacement operator ${\hat{{\bf{R}}}}\left({\bf{r}}\right)$ as in \cite{cite_33}
\begin{eqnarray} 
{\hat{{\bf{R}}}}\left({\bf{r}}\tau\right)&=&\frac{1}{\beta{N_{\rm ph}}}\sum_{{\bf{q}}\Omega_{m}}\sqrt{\frac{\hbar}{{2M}^{\ast}\Omega_{{\bf{q}}}}}{\bf{e}}\left({\bf{q}}\right)\left[c_{{\bf{q}}}\left(\Omega_{m}\right)e^{i\left({\bf{q}}{\bf{r}}-\Omega_{m}\tau\right)}\right.
\nonumber\\
&&\left.+c^{\dag}_{{\bf{q}}}\left(\Omega_{m}\right)e^{-i\left({\bf{q}}{\bf{r}}-\Omega_{m}\tau\right)}\right],
\label{Equation_6}
\end{eqnarray}
where $\Omega_{m}=2\pi{m}/\beta$ are bosonic Matsubara frequencies (with $m=0,\pm{1}, \pm{2}, ...$), ${\bf{e}}\left({\bf{q}}\right)$ is the unit vector in the direction of the wave vector ${\bf{q}}$ and $M^{\ast}=M/S_{x}S_{y}$ (with $S_{i}=\left(2\pi\right)^{2}$) being the renormalized mass for the single phonon excitation. Next, $N_{\rm ph}$ is the number of phonon modes in the system and the operators $c^{\dag}_{{\bf{q}}}\left(\Omega_{m}\right)$ and $c_{{\bf{q}}}\left(\Omega_{m}\right)$, in Eq.(\ref{Equation_6}), are the operators of phonon creation and annihilation with the wave vector ${\bf{q}}$ and frequency $\Omega_{m}$. 
%
%
%
%
The operator $\delta{\hat{E}}$, in Eqs.(\ref{Equation_4}) and (\ref{Equation_5}), is, indeed, Hamiltonian operator of electron-phonon interaction in the system, because it gives the total electron energy modification in the system caused by the presence of small phononic displacement ${\hat{{\bf{R}}}}\left({\bf{r}}\tau\right)$. We can write
\begin{eqnarray} 
{\cal{H}}_{\rm int}=-ea^{2}C\int{d{\bf{r}}\hat{n}\left({\bf{r}}\right){\rm div}{\hat{{\bf{R}}}}\left({\bf{r}}\right)},
\label{Equation_7}
\end{eqnarray}
where $C=ZeN_{\rm ion}/\upsilon$ (with $N_{\rm ion}$ is the number of carbon atoms in graphene lattice and $\upsilon$ being the volume of the sample). For the total electron concentration $\hat{n}\left({\bf{r}}\right)$, in Eq.(\ref{Equation_7}), we have in our case 
\begin{eqnarray} 
\hat{n}\left({\bf{r}}\right)=\hat{n}_{a}\left({\bf{r}}\right)+\hat{n}_{b}\left({\bf{r}}\right).
\label{Equation_8}
\end{eqnarray}
The next Section is devoted to the discussion about single-particle ${\bf{q}}$-excitations, which are responsible for the SC state in the system.
%
\subsection{\label{sec:Section_2_2} Single-particle ${\bf{q}}$-excitations and phonon relaxation time}
%
We consider two types of single-quasiparticle excitations for the electrons, namely, those with ${\bf{k}}_{1}={\bf{k}}+\frac{{\bf{q}}}{2}$, $\nu_{1n}=\nu_{n}-\frac{\Omega_{m}}{2}$ and ${\bf{k}}_{2}={\bf{k}}-\frac{{\bf{q}}}{2}$, $\nu_{2n}=\nu_{n}+\frac{\Omega_{m}}{2}$, where $\nu_{n}=\pi{\left(2n+1\right)}/\beta$ (with $n=0,\pm 1, \pm 2, ...$) are fermionic Matsubara frequencies and the ${\bf{q}}$ wave vector exactly coincides with that of the single-phonon excitation vector, given in Eq.(\ref{Equation_6}). 

\begin{figure}
	\includegraphics[scale=1.1]{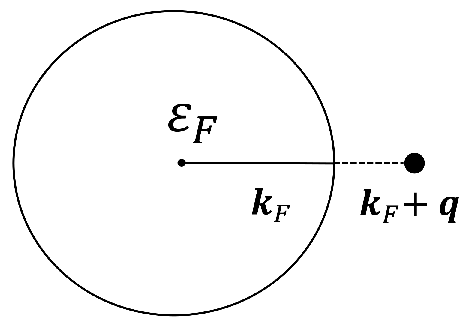}
	\caption{\label{fig:Fig_2}(Color online) Single-particle ${\bf{q}}$-excitation above Fermi surface with a wave vector ${\bf{q}}$, along Fermi vector ${\bf{k}}_{\rm F}$.}
\end{figure}

\begin{figure}
	\includegraphics[scale=0.65]{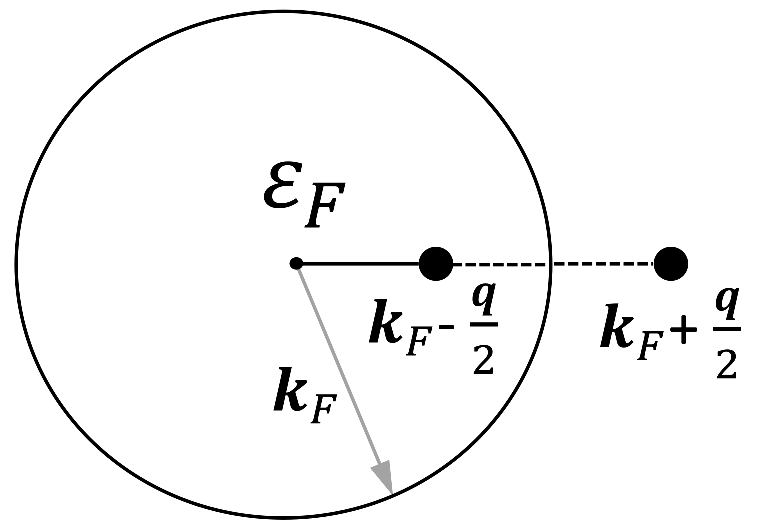}
	\caption{\label{fig:Fig_3}(Color online) Single-particle ${\bf{q}}$-excitation discretization above and below Fermi surface with a wave vector ${\bf{q}}/2$, according to discussion in Eq.(\ref{Equation_8}).}
\end{figure}

We will write the fermionic quasiparticles in terms of Grassmann complex variables \cite{cite_34}.
Then, after putting the expression of ${\hat{{\bf{R}}}}\left({\bf{r}}\tau\right)$, in Eq.(\ref{Equation_6}), into Eq.(\ref{Equation_7}), we get action of the electron-phonon interaction in the Fourier space representation
\begin{widetext}
\begin{eqnarray}
{\cal{S}}_{\rm int}=\int^{\beta}_{0}d\tau{\cal{H}}_{\rm int}\left(\tau\right)&=&\frac{1}{\beta^{2}NN_{\rm ph}}\sum_{\substack{{\bf{k}}{\bf{q}},\sigma\\ \nu_{n},\Omega_{m}}}\sum_{\eta=a,b}i\alpha_{{\bf{q}}}\left(\bar{\eta}_{{\bf{k}}+\frac{{\bf{q}}}{2},\sigma}\left(\nu_{n}-\frac{\Omega_{m}}{2}\right){\eta}_{{\bf{k}}-\frac{{\bf{q}}}{2},\sigma}\left(\nu_{n}+\frac{\Omega_{m}}{2}\right)\right)\hat{c}_{{\bf{q}}}\left(\Omega_{m}\right)
\nonumber\\
&&-i\alpha_{{\bf{q}}}\left(\bar{\eta}_{{\bf{k}}-\frac{{\bf{q}}}{2},\sigma}\left(\nu_{n}+\frac{\Omega_{m}}{2}\right){\eta}_{{\bf{k}}+\frac{{\bf{q}}}{2},\sigma}\left(\nu_{n}-\frac{\Omega_{m}}{2}\right)\right)\bar{\hat{c}}_{{\bf{q}}}\left(\Omega_{m}\right),
\label{Equation_9}
\end{eqnarray}
\end{widetext}
where $\alpha_{{\bf{q}}}$, in Eq.(\ref{Equation_9}), is the energy of the electron-phonon interaction
\begin{eqnarray}
\alpha_{{\bf{q}}}=\sqrt{\frac{\hbar{ea^{2}C}}{2m\Omega_{q}}}|{\bf{q}}|.
\label{Equation_10}
\end{eqnarray}
Thus, we have a single quasiparticle excitation with a small wave vector ${\bf{q}}$ above Fermi surface (see in Fig.~\ref{fig:Fig_2}) which coincides with that of the phonon wave vector excited by the external light source \cite{cite_35} (see in Fig.~\ref{fig:Fig_1}). We represented it in the form of a product of two single-particle excitations with the wave vectors ${\bf{k}}+{\bf{q}}/2$ and ${\bf{k}}-{\bf{q}}/2$ (see in Fig.~\ref{fig:Fig_3}). Indeed, when exciting a single quasiparticle above Fermi surface with a small phonon wave vector ${\bf{q}}$, the energy (per unit volume of the sample), necessary for the excitation is
\begin{eqnarray} 
\Delta{\varepsilon}\approx\frac{1}{\upsilon}\left(\varepsilon\left({\bf{k}}_{F}+{\bf{q}}\right)-\varepsilon\left({\bf{k}}_{F}\right)\right)=\rho\left(\varepsilon_{F}\right)\varepsilon^{\left({\bf{q}}\right)}_{1}\varepsilon^{\left({\bf{q}}\right)}_{2},
\label{Equation_11}
\end{eqnarray}  
where ${\bf{k}}_{\rm F}$ is Fermi wave vector. Thus, we ascribe $\varepsilon^{\left({\bf{q}}\right)}_{1}$ and $\varepsilon^{\left({\bf{q}}\right)}_{2}$ the single-quasiparticle excitations and the product of those excitation energies is multiplied with the density of states at Fermi level $\rho\left(\varepsilon_{F}\right)=mk_{F}/\pi^{2}\hbar^{2}$. For a single excitation, we have
%
%
%
%
\begin{eqnarray}
\varepsilon^{\left({\bf{q}}\right)}_{1}=\varepsilon^{\left({\bf{q}}\right)}_{2}\equiv\varepsilon_{\rm exc}=\frac{\pi\hbar^{2}}{m}\sqrt{\frac{q}{\upsilon}}=\frac{\hbar^{2}{\bf{q}}'^{2}}{2m^{\ast}_{\rm el}}.
\label{Equation_12}
\end{eqnarray}
We putted ${\bf{q}}'^{2}=\sqrt{|{\bf{q}}|/\upsilon}$ (or ${\bf{q}}'=\pm\sqrt[4]{{\bf{q}}/\upsilon}$) and $m^{\ast}_{\rm el}=m_{\rm el}/2\pi$.
Or, equivalently, in energy units, we can write 
\begin{eqnarray}
\varepsilon_{\rm exc}=\frac{{\pi}e^{4}}{2}\sqrt{\frac{q}{\upsilon}}{\rm {Ry}}^{-1},
\label{Equation_13}
\end{eqnarray}  
where ${\rm {Ry}}=me^{4}/2\hbar^{2}=13.6$ eV is the Rydberg constant. 
Throughout the paper, we use the units in which $e\equiv 1$ and $\upsilon\equiv 1$. We consider, here, a small excitation around Fermi level in graphene with the excitation wave vector ${\bf{q}}=0.1{{\bf{K}}}$. In this case, we get for two-quasiparticle ${\bf{q}}$-excitation energy (above or below Fermi level) 
%
%
%
%
\begin{eqnarray} 
\varepsilon_{\rm exc}=0.0567  \ \ {\rm eV} 
\label{Equation_14}
\end{eqnarray}  
and we will normalize band-energy parameters in the system to $\varepsilon_{\rm exc}$, as the unit of measure for the energy scales, while the externally controllable parameters which affect directly the lattice of graphene will be normalized to the hopping amplitude $\gamma_0$ of the electrons in graphene. As we will see, later on, in Section \ref{sec:Section_4}, the value in Eq.(\ref{Equation_14}) has a strong relation with our further numerical evaluations. The obtained quantized excitations in Eq.(\ref{Equation_12}) could be regarded as a discretization (see in Fig.~\ref{fig:Fig_3}) of the possible single-particle ${\bf{q}}$-excitation in the system.

Indeed, when exciting the system with an external wave, with the wave vector ${\bf{Q}}$, the spring force appears which acts on the atoms relocated from their equilibrium positions. In order to have a continuous wave propagation in the system, with the wave vector coinciding with that of the phononic wave vector ${\bf{q}}={\bf{Q}}$, we need a certain relaxation time $\tau_{\rm rel}$ \cite{cite_35, cite_36, cite_37, cite_38, cite_39}. It could be calculated after equating the change in atomic impulsion $\Delta{P}$, during that time, with the impulsion of the external excitation
\begin{eqnarray}      
\Delta{P}=\Delta{F}\Delta{\tau}_{rel}=\hbar\left|{\bf{Q}}\right|,
\label{Equation_15}
\end{eqnarray}
where $\Delta{F}=\kappa_{\rm sp}\Delta{R}$ is the spring force, and $\Delta{R}=\left\langle R\right\rangle$ is the average deviation of the atoms from their equilibrium positions, or, in other words, the phononic mean free path. Next, $\kappa_{\rm sp}$, in Eq.(\ref{Equation_15}), is the spring constant, which, in our case of graphene, is of order $\kappa_{\rm sp}=0.01-0.05$ N/cm \cite{cite_40}, and $\tau_{\rm rel}$ is the lattice relaxation time. 
The average $\left\langle R \right\rangle$ can be calculated by following formula
\begin{eqnarray} 
\left\langle R \right\rangle=\frac{\int^{R_{\rm max}}_{0}dR R e^{-\frac{\Delta{F}R}{k_{B}T}}}{\int^{R_{\rm max}}_{0}e^{-\frac{\Delta{F}R}{k_{B}T}}dR}.
\label{Equation_16}
\end{eqnarray}
The calculation of the right-hand side in Eq.(\ref{Equation_16}) gives the following self-consistent rule for the mean free path $\left\langle R \right\rangle$
\begin{eqnarray}
\left\langle R \right\rangle=\frac{k_{\rm B}T}{\Delta{F}}\left(1-\frac{\Delta{F}{R_{\rm max}}}{k_{\rm B}T}n_{\rm B}\left(\frac{\Delta{F}{R_{\rm max}}}{k_{\rm B}T}\right)\right),
\label{Equation_17}
\end{eqnarray}
where $n_{\rm B}\left(x\right)$ is Bose-Einstein distribution function
\begin{eqnarray}
n_{\rm B}\left(x\right)=\frac{1}{e^{\beta{x}}-1}.
\label{Equation_18}
\end{eqnarray}
We can calculate the phonon relaxation time for the limiting cases. Namely, for $\Delta{F}{R_{\rm max}}\gg k_{B}T$, we get for the average displacement $\left\langle R\right\rangle\sim \sqrt{k_{B}T/\kappa_{\rm sp}}$, and the relaxation time is of order of
\begin{eqnarray}
\tau_{\rm rel}\sim\frac{\hbar|\bf{Q}|}{\sqrt{\kappa{k_{\rm B}}}}\frac{1}{\sqrt{T}}.
\label{Equation_19}
\end{eqnarray}
For, example, when exciting the system at Dirac wave vector ${\bf{Q}}={\bf{K}}$ (as it has been done in Ref.\onlinecite{cite_35}, using the constrained density functional perturbation theory), the time necessary for attenuation of the phononic wave with the frequency $\Omega_{{\bf{q}}={\bf{Q}}={\bf{K}}}$ is of order $\tau_{\rm rel}\sim 10.1$ ps. (see in Eq.(\ref{Equation_19})), which agrees well with the recently obtained results in Ref.\onlinecite{cite_35, cite_36}, where a phonon relaxation time scale was obtained in the range of $1-10$ ps, concerning the ultrafast electron-phonon relaxation dynamics in graphene and the formation of the ordered states and phase transitions induced by the external photoexcitation \cite{cite_35}.   
%
                                                                       
\section{\label{sec:Section_3} Phonon field integration}
%
\subsection{\label{sec:Section_3_1} The effective fermionic action}
%
In this Section, we define the following quantities 
\begin{widetext}
\begin{eqnarray}
\Lambda_{{\bf{q}}}\left(\Omega_{m}\right)=\frac{g^{\ast}_{{\bf{q}}}}{\beta{N}}\sum_{{\bf{k}}\nu_{n},\sigma}\sum_{\eta=a,b}\left[\bar{\eta}_{{\bf{k}}-\frac{{\bf{q}}}{2},\sigma}\left(\nu_{n}+\Omega_{m}/{2}\right){\eta}_{{\bf{k}}+\frac{{\bf{q}}}{2},\sigma}\left(\nu_{n}-\Omega_{m}/{2}\right)\right],
\nonumber\\
\bar{\Lambda}_{{\bf{q}}}\left(\Omega_{m}\right)=\frac{g_{{\bf{q}}}}{\beta{N}}\sum_{{\bf{k}}\nu_{n},\sigma}\sum_{\eta=a,b}\left[\bar{\eta}_{{\bf{k}}+\frac{{\bf{q}}}{2},\sigma}\left(\nu_{n}-\Omega_{m}/{2}\right){\eta}_{{\bf{k}}-\frac{{\bf{q}}}{2},\sigma}\left(\nu_{n}+\Omega_{m}/{2}\right)\right],
\label{Equation_20}
\end{eqnarray} 
\end{widetext}
where $g_{{\bf{q}}}=i\alpha_{{\bf{q}}}$. Then, the action corresponding to the electron-phonon interaction in Eq.(\ref{Equation_9}) could be written as
\begin{eqnarray}
{\cal{S}}_{\rm int}=\frac{1}{\beta{N}_{\rm ph}}\sum_{{\bf{q}}\Omega_{m}}\left[\bar{\Lambda}_{{\bf{q}}}\left(\omega_{m}\right)c_{{\bf{q}}}\left(\Omega_{m}\right)\right.
\nonumber\\
\left.+\bar{c}_{{\bf{q}}}\left(\Omega_{m}\right){\Lambda}_{{\bf{q}}}\left(\Omega_{m}\right)\right].
\label{Equation_21}
\end{eqnarray} 
The total action of graphene system reads as
\begin{eqnarray}
{\cal{S}}_{\rm gr}={\cal{S}}_{\rm t-B}+{\cal{S}}_{\rm U-V}+{\cal{S}}_{\rm ph}+{\cal{S}}^{\rm ph}_{\rm B}+{\cal{S}}^{\rm el}_{\rm B}+{\cal{S}}_{\rm int},
\label{Equation_22}
\end{eqnarray} 
where
\begin{eqnarray}
&&{\cal{S}}_{\rm t-B}=\int^{\beta}_{0}d\tau \hat{\cal{H}}_{\rm t-B}\left(\tau\right),
\nonumber\\
&&{\cal{S}}_{\rm U-V}=\int^{\beta}_{0}d\tau \hat{\cal{H}}_{\rm U-V}\left(\tau\right),
\nonumber\\
&&{\cal{S}}_{\rm ph}=\int^{\beta}_{0}d\tau \hat{\cal{H}}_{\rm ph}\left(\tau\right)
\nonumber\\
&&=\frac{1}{\beta{N}_{\rm ph}}\sum_{{\bf{q}}\Omega_{m}}\hbar{\Omega_{q}}\bar{c}_{\bf{q}}\left(\Omega_{m}\right)c_{\bf{q}}\left(\Omega_{m}\right),
\nonumber\\
&&{\cal{S}}^{\rm ph}_{\rm B}=\frac{1}{\beta{N}_{\rm ph}}\sum_{{\bf{q}}\Omega_{m}}\bar{c}_{{\bf{q}}}\left(\Omega_{m}\right)\left(-i\Omega_{m}\right){c}_{{\bf{q}}}\left(\Omega_{m}\right),
\nonumber\\
&&{\cal{S}}^{\rm el}_{\rm B}=\frac{1}{\beta{N}}\sum_{{\bf{k}}\nu_{n}}\sum_{\eta=a,b}\bar{\eta}_{{\bf{k}}\sigma}\left(\nu_{n}\right)\left(-i\nu_{n}\right){\eta}_{{\bf{k}}\sigma}\left(\nu_{n}\right).
\label{Equation_23}
\end{eqnarray} 
The last two terms in Eq.(\ref{Equation_17}) are Berry actions for phonons and electrons, respectively.  
Next, we integrate out the phonons via Hubbard-Stratonovich transformation rule
\begin{eqnarray}
&&\int\left[{\cal{D}}\bar{c}{\cal{D}}c\right]\prod_{\eta=a,b}\exp\left(-\frac{{\cal{S}}_{\rm ph}\left[\bar{c},c\right]}{2}-\frac{{\cal{S}}^{\rm ph}_{\rm B}\left[\bar{c},c\right]}{2}-\right.
\nonumber\\
&&\left.-{\cal{S}}_{\rm int}\left[\bar{c},c,\bar{\eta},\eta\right]\right)\cong \exp\left(-\frac{1}{\beta{N}_{\rm ph}}\sum_{{\bf{q}}\Omega_{m}}\frac{|\Lambda_{{\bf{q}}}\left(\Omega_{m}\right)|^{2}}{i\Omega_{m}+\hbar{\Omega_{\bf{q}}}}\right).
\nonumber\\
\label{Equation_24}
\end{eqnarray}

The summation of type $\sum_{\Omega_{m}}|\Lambda_{{\bf{q}}}\left(\Omega_{m}\right)|^{2}\Omega_{m}/\left(\left(\hbar{\Omega_{{\bf{q}}}}\right)^{2}+\Omega^{2}_{m}\right)$ gives zero due to the parity with respect to $\Omega_{m}$ of the four fermion product $|\Lambda_{{\bf{q}}}\left(\Omega_{m}\right)|^{2}$. Therefore, we get the effective fermionic action in the form 
\begin{eqnarray}
{\cal{S}}_{\rm eff}=\frac{1}{\beta{N}_{\rm ph}}\sum_{{\bf{q}}\Omega_{m}}\frac{\hbar\Omega_{\bf{q}}}{\left(\hbar{\Omega_{{\bf{q}}}}\right)^{2}+\Omega^{2}_{m}}|\Lambda_{\bf{q}}\left(\Omega_{m}\right)|^{2},
\label{Equation_25}
\end{eqnarray}
and the total action of graphene in Eq.(\ref{Equation_22}) will be reformed to the form
\begin{eqnarray}
{\cal{S}}_{\rm gr}={\cal{S}}_{\rm t-B}+{\cal{S}}_{\rm U-V}+{\cal{S}}^{\rm el}_{\rm B}+{\cal{S}}_{\rm eff}.
\label{Equation_26}
\end{eqnarray}
Furthermore, we have 
\begin{widetext}
\begin{eqnarray}
|\Lambda_{\bf{q}}\left(\Omega_{m}\right)|^{2}=\frac{\alpha^{2}_{{\bf{q}}}}{\beta^{2}N^{2}}\sum_{\substack{{\bf{k}}_{1}{\bf{k}}_{2}\\ \nu_{n}{\nu'_{n}}}}\sum_{\substack{\eta\eta'\\ \sigma\sigma'}}\bar{\eta}_{{\bf{k}}_{1}+\frac{{\bf{q}}}{2},\sigma}\left(\nu_{n}-\Omega_{m}/2\right){\eta}_{{\bf{k}}_{1}-\frac{{\bf{q}}}{2},\sigma}\left(\nu_{n}+\Omega_{m}/2\right)\bar{\eta}'_{{\bf{k}}_{2}-\frac{{\bf{q}}}{2},\sigma'}\left(\nu'_{n}+\Omega_{m}/2\right){\eta}'_{{\bf{k}}_{2}+\frac{{\bf{q}}}{2},\sigma'}\left(\nu'_{n}-\Omega_{m}/2\right).
\nonumber\\
\label{Equation_27}
\end{eqnarray}
\end{widetext}
Hereafter, we consider the case of ${\bf{q}}$-excitations on the uniform background of ${\bf{k}}$-space, therefore we put ${\bf{k}}_{1}={\bf{k}}_{2}$, $\nu_{n}=\nu'_{n}$, $\sigma=\sigma'$ (for the triplet superconducting pairing) and $\eta=\eta'$ (remember that standard considerations in Bardeen-Cooper-Schrieffer theory of superconductivity concern the case ${\bf{k}}_{2}=-{\bf{k}}_{1}$ and ${\bf{q}}=0$). The condition $\eta=\eta'$, here, assumes that the superconductivity occurs locally, i.e., at the sublattice sites $a$ or $b$, and there is no pairing between the electrons on different lattice sites.  
%
\subsection{\label{sec:Section_3_2} Linearization of Hubbard-$U$ term}
%
The first term, in Eq.(\ref{Equation_3}), is quadratic in terms of the electron density operator $\hat{n}\left({\bf{r}}\right)$. It could be linearized via Hubbard-Stratonovich transformation procedure. Indeed, after passing to complex Grassmann representation, we have
\begin{eqnarray}
Un_{\eta\uparrow}\left({\bf{r}}\tau\right)n_{\eta\downarrow}\left({\bf{r}}\tau\right)=\frac{U}{4}\left(n^{2}_{\eta}\left({\bf{r}}\tau\right)-p^{2}_{{\rm z} \eta}\left({\bf{r}}\tau\right)\right),
\label{Equation_28}
\end{eqnarray}
where $n_{\eta}\left({\bf{r}}\tau\right)=n_{\eta\uparrow}\left({\bf{r}}\tau\right)+n_{\eta\downarrow}\left({\bf{r}}\tau\right)$ is the total electron density and $p_{{\rm z}\eta}\left({\bf{r}}\tau\right)$ is electron density polarization $p_{{\rm z}\eta}\left({\bf{r}}\tau\right)=n_{\eta\uparrow}\left({\bf{r}}\tau\right)-n_{\eta\downarrow}\left({\bf{r}}\tau\right)$. Decoupling procedure is following
\begin{eqnarray}
&&\exp\left(\int^{\beta}_{0}d\tau\sum_{{\bf{r}}}in_{\eta}\left({\bf{r}}\tau\right)\frac{U}{4}in_{\eta}\left({\bf{r}}\tau\right)\right)=
\nonumber\\
&&=\int{\left[{\cal{D}}V_{\eta}\right]}e^{-\frac{1}{2U}\int^{\beta}_{0}d\tau\sum_{{\bf{r}}}V^{2}_{\eta}\left({\bf{r}}\tau\right)+\int^{\beta}_{0}d\tau\sum_{{\bf{r}}}in_{\eta}\left({\bf{r}}\tau\right)V_{\eta}\left({\bf{r}}\tau\right)},
\nonumber\\
\label{Equation_29}
\end{eqnarray}
where $V_{\eta}\left({\bf{r}}\tau\right)$ is the auxiliary decoupling potential. Furthermore, evaluating the integral by the saddle-point method, we get the contribution to the total action coming from Hubbard-$U$ term. 
\begin{eqnarray}
{\cal{S}}_{\rm U}=-\frac{U}{2}\int^{\beta}_{0}d\tau\sum_{{\bf{r}}\eta}n_{\eta}\left({\bf{r}}\tau\right)\left\langle n_{\eta}\left({\bf{r}}\tau\right)\right\rangle.
\label{Equation_30}
\end{eqnarray}
This action is already linear in electron density term. The average $\left\langle n\left({\bf{r}}\tau\right)\right\rangle$ could be evaluated with the help of the partition function
\begin{eqnarray}
{\cal{Z}}_{\rm gr}=\int{\left[{\cal{D}}\bar{a}{\cal{D}}a\right]}\int{\left[{\cal{D}}\bar{b}{\cal{D}}b\right]}e^{-{\cal{S}}_{\rm gr}\left[\bar{a},a,{\bar{b}},b\right]}.
\label{Equation_31}
\end{eqnarray}
We have 

\begin{eqnarray}
\left\langle n_{\eta}\left({\bf{r}}\tau\right)\right\rangle=\frac{\int{\left[{\cal{D}}\bar{a}{\cal{D}}a\right]}\int{\left[{\cal{D}}\bar{b}{\cal{D}}b\right]}n_{\eta}\left({\bf{r}}\tau\right)e^{-{\cal{S}}_{\rm gr}\left[\bar{a},a;{\bar{b}},b\right]}}{\int{\left[{\cal{D}}\bar{a}{\cal{D}}a\right]}\int{\left[{\cal{D}}\bar{b}{\cal{D}}b\right]}e^{-{\cal{S}}_{\rm gr}\left[\bar{a},a;{\bar{b}},b\right]}}.
\nonumber\\
\label{Equation_32}
\end{eqnarray}
The same procedure could be repeated also for the polarization term in Eq.(\ref{Equation_28}), however, for present, we suppose that the resulting antiferromagnetic order parameter is zero, i.e., $\Delta_{{\rm AFM}\eta}=U\left\langle p_{{\rm z}\eta}\left({\bf{r}}\tau\right)\right\rangle/2=0$, and it follows that $\left\langle n_{\eta\uparrow}\left({\bf{r}}\tau\right)\right\rangle=\left\langle n_{\eta\downarrow}\left({\bf{r}}\tau\right)\right\rangle$.  

\subsection{\label{sec:Section_3_3} Nambu spinors and Green's function matrix}
%
The action in Eq.(\ref{Equation_25}) could be simplified by supposing the small ${\bf{q}}$-excitation energies. Namely we have 
\begin{widetext}
\begin{eqnarray}
{\cal{S}}_{\rm eff}&=&-\frac{\pi^{-1}}{\beta^{2}{N}N_{\rm ph}}\sum_{\substack{{\bf{k}}{\bf{q}}\\ \nu_{n}\Omega'_{m}}}\sum_{\substack{\eta=a,b\\ \sigma}}\frac{g'_{{\bf{q}}}\omega_{\bf{q}}}{\omega^{2}_{\bf{q}}+\left(\Omega'_{m}\right)^{2}}\bar{\eta}_{{\bf{k}}+\frac{{\bf{q}}}{2},\sigma}\left(\nu_{n}-\Omega'_{m}/2\right)\bar{\eta}_{{\bf{k}}-\frac{{\bf{q}}}{2},\sigma}\left(\nu_{n}+\Omega'_{m}/2\right)\times
\nonumber\\
&&\times{\eta}_{{\bf{k}}-\frac{{\bf{q}}}{2},\sigma}\left(\nu_{n}+\Omega'_{m}/2\right){\eta}_{{\bf{k}}_{2}+\frac{{\bf{q}}}{2},\sigma}\left(\nu_{n}-\Omega'_{m}/2\right),
\label{Equation_33}
\end{eqnarray}   
\end{widetext}
where $g'_{{\bf{q}}}=\pi\alpha^{2}_{{\bf{q}}}\beta$ is the renormalized electron-phonon interaction energy and the dimensionless parameters $\omega_{\bf{q}}=\beta\hbar\Omega_{\bf{q}}$ and $\Omega'_{m}=\beta\Omega_{m}$ have been introduced. Furthermore, we use the fact that the excitation energy $\varepsilon_{{\bf{q}}}=\hbar\Omega_{\bf{q}}$ is small, which permits to transform the effective electron-phonon interaction potential to a Lorentzian form:
\begin{eqnarray}
V_{\bf{q}}=\frac{1}{\pi}\frac{g'_{{\bf{q}}}\omega_{\bf{q}}}{\omega^{2}_{\bf{q}}+\left(\Omega'_{m}\right)^{2}}.
\label{Equation_34}
\end{eqnarray} 
In this way, we simplify the expression in Eq.(\ref{Equation_25}) and only the mode $\Omega'_{m}=0$ survives from the summation over Bosonic frequencies. Moreover, in the limit of non-dressed phonons ($\hbar\Omega_{\bf{q}}=\hbar^{2}{\bf{q}}^{2}/2m$) the electron-phonon coupling energy $g'_{{\bf{q}}}$ do not depends on the wave-vector ${\bf{q}}$. Indeed, from Eq.(\ref{Equation_10}), we have $\alpha_{{\bf{q}}}=|a|\sqrt{eC}\equiv\lambda_{\rm eff}$ and 
\begin{eqnarray}
g'_{\bf{q}}\equiv g=\pi\lambda^{2}_{\rm eff}\beta. 
\label{Equation_35}
\end{eqnarray} 
Next, we define the triplet superconducting order parameters $\Delta^{\rm sc}_{\rm \eta\sigma}$ and $\bar{\Delta}^{\rm sc}_{\eta\sigma}$ 
\begin{eqnarray}
\Delta^{\rm sc}_{\rm \eta,\sigma}=\frac{2g}{\beta^{2}{N}^{2}N_{\rm ph}}\sum_{{\bf{k}}{\bf{q}},\nu_{n}}\left\langle \eta_{{\bf{k}}-\frac{{\bf{q}}}{2},\sigma}\left(\nu_{n}\right)\eta_{{\bf{k}}+\frac{{\bf{q}}}{2},\sigma}\left(\nu_{n}\right)\right\rangle,
\nonumber\\
\bar{\Delta}^{\rm sc}_{\rm \eta,\sigma}=\frac{2g}{\beta^{2}{N}^{2}N_{\rm ph}}\sum_{{\bf{k}}{\bf{q}},\nu_{n}}\left\langle \bar{\eta}_{{\bf{k}}+\frac{{\bf{q}}}{2},\sigma}\left(\nu_{n}\right)\bar{\eta}_{{\bf{k}}-\frac{{\bf{q}}}{2},\sigma}\left(\nu_{n}\right)\right\rangle.
\label{Equation_36}
\end{eqnarray} 
We suppose that $\Delta^{\rm sc}_{\rm a\sigma}=\Delta^{\rm sc}_{\rm b\sigma}\equiv\Delta^{\rm sc}_{\sigma}$, which corresponds to the homogeneous distribution of possible superconducting order in graphene.
Then, the action in Eq.(\ref{Equation_33}) will be written in the following form
\begin{eqnarray}
{\cal{S}}_{\rm eff}&=&-\frac{g}{\beta{N}N_{\rm ph}}\sum_{{\bf{k}}{\bf{q}}}\sum_{\nu_{n}}\sum_{\eta\sigma}|A^{\left(\eta\right)}_{{\bf{k}}{\bf{q}},\sigma}\left(\nu_{n}\right)|^{2},
\label{Equation_37}
\end{eqnarray}
where 
\begin{eqnarray}
A^{\left(\eta\right)}_{{\bf{k}}{\bf{q}},\sigma}\left(\nu_{n}\right)={\eta}_{{\bf{k}}-\frac{{\bf{q}}}{2},\sigma}\left(\nu_{n}\right){\eta}_{{\bf{k}}+\frac{{\bf{q}}}{2}\sigma}\left(\nu_{n}\right),
\nonumber\\
\bar{A}^{\left(\eta\right)}_{{\bf{k}}{\bf{q}},\sigma}\left(\nu_{n}\right)=\bar{\eta}_{{\bf{k}}+\frac{{\bf{q}}}{2},\sigma}\left(\nu_{n}\right)\bar{\eta}_{{\bf{k}}-\frac{{\bf{q}}}{2},\sigma}\left(\nu_{n}\right).
\label{Equation_38}
\end{eqnarray}
Furthermore, the expression under sums in the right-hand side in Eq.(\ref{Equation_37}) is biquadratic with respect to fermionic variables and, therefore, is suitable for further Hubbard-Stratonovich decoupling procedure. Indeed, we have
\begin{eqnarray}
&&\exp\left(\frac{g}{\beta{N}N_{\rm ph}}\sum_{{\bf{k}}{\bf{q}},\sigma}\sum_{\nu_{n}}|A^{\left(\eta\right)}_{{\bf{k}}{\bf{q}},\sigma}\left(\nu_{n}\right)|^{2}\right)=
\nonumber\\
&&=\int{\left[{\cal{D}}\bar{\zeta}{\cal{D}}\zeta\right]}\exp\left(\frac{1}{\beta{N}N_{\rm ph}}\sum_{{\bf{k}}{\bf{q}},\sigma}\sum_{\nu_{n}}-\frac{1}{g}|\zeta_{{\bf{k}}{\bf{q}},\sigma}\left(\nu_{n}\right)|^{2}\right.
\nonumber\\
&&\left.+\bar{\zeta}_{{\bf{k}}{\bf{q}},\sigma}\left(\nu_{n}\right)A^{\left(\eta\right)}_{{\bf{k}}{\bf{q}},\sigma}\left(\nu_{n}\right)+\bar{A}^{\left(\eta\right)}_{{\bf{k}}{\bf{q}},\sigma}\left(\nu_{n}\right){\zeta}_{{\bf{k}}{\bf{q}},\sigma}\left(\nu_{n}\right)\frac{}{}\right),
\label{Equation_39}
\end{eqnarray}
where $\bar{\zeta}_{{\bf{k}}{\bf{q}},\sigma}\left(\nu_{n}\right)$ and $\zeta_{{\bf{k}}{\bf{q}},\sigma}\left(\nu_{n}\right)$ are the auxiliary source fields. 
The exponential, in the right-hand side in Eq.(\ref{Equation_39}), for a given mode $\left({\bf{k}},{\bf{q}};\nu_{n}\right)$, could be regarded as a generating functional of variables $\bar{\zeta}$ and $\zeta$, i.e., 
\begin{eqnarray}
&&{\cal{G}}\left[\bar{\zeta},\zeta\right]=\exp\left(-\frac{1}{g}|\zeta_{{\bf{k}}{\bf{q}},\sigma}\left(\nu_{n}\right)|^{2}+\right.
\nonumber\\
&&\left.+\bar{\zeta}_{{\bf{k}}{\bf{q}},\sigma}\left(\nu_{n}\right)A^{\left(\eta\right)}_{{\bf{k}}{\bf{q}},\sigma}\left(\nu_{n}\right)+\bar{A}^{\left(\eta\right)}_{{\bf{k}}{\bf{q}},\sigma}\left(\nu_{n}\right){\zeta}_{{\bf{k}}{\bf{q}},\sigma}\left(\nu_{n}\right)\frac{}{}\right).
\nonumber\\
\label{Equation_40}
\end{eqnarray}
The functional differentiation of Eq.(\ref{Equation_40}), with respect to $\bar{\zeta}$ and $\zeta$, gives the saddle-point values. For example, the equation for $\zeta$ reads
\begin{eqnarray}
&&\frac{\delta}{\delta{\bar{\zeta}}}\int{\left[{\cal{D}}\bar{\zeta}{\cal{D}}\zeta\right]}{\cal{G}}\left[\bar{\zeta},\zeta\right]=
\nonumber\\
&&=\int{\left[{\cal{D}}\bar{\zeta}{\cal{D}}\zeta\right]}\frac{\delta}{\delta{\bar{\zeta}}}\left({\cal{G}}\left[\bar{\zeta},\zeta\right]\right)=
\nonumber\\
&&=\int{\left[{\cal{D}}\bar{\zeta}{\cal{D}}\zeta\right]}\left(-\frac{1}{g}\zeta_{{\bf{k}}{\bf{q}},\sigma}\left(\nu_{n}\right)+A^{\left(\eta\right)}_{{\bf{k}}{\bf{q}},\sigma}\left(\nu_{n}\right)\right){\cal{G}}\left[\bar{\zeta},\zeta\right]=
\nonumber\\
&&=0,
\label{Equation_41}
\end{eqnarray}
and we obtain for the saddle-point value
\begin{eqnarray}
\zeta^{\rm s.p.}_{\sigma}=g\left\langle A^{\left(\eta\right)}_{{\bf{k}}{\bf{q}},\sigma}\left(\nu_{n}\right) \right\rangle. 
\label{Equation_42}
\end{eqnarray}
Similarly, we find for the $\bar{\zeta}^{\rm s.p.}$
\begin{eqnarray}
\bar{\zeta}^{\rm s.p.}_{\sigma}=g\left\langle \bar{A}^{\left(\eta\right)}_{{\bf{k}}{\bf{q}},\sigma}\left(\nu_{n}\right) \right\rangle. 
\label{Equation_43}
\end{eqnarray} 
The contribution to the fermionic action, coming from above-mentioned decoupling procedure, is 
\begin{eqnarray}
&&{\cal{S}}_{\rm sc}\left[\bar{\Delta}^{\rm sc}, \Delta^{\rm sc}\right]=
\nonumber\\
&&=-\frac{1}{2\beta{N}N_{\rm ph}}\sum_{\substack{{\bf{k}}{\bf{q}}\\ \nu_{n}\sigma}}\left(\bar{\Delta}^{\rm sc}_{\sigma}\eta_{{\bf{k}}-\frac{\bf{q}}{2},\sigma}\left(\nu_{n}\right)\eta_{{\bf{k}}+\frac{\bf{q}}{2},\sigma}\left(\nu_{n}\right)+\right.
\nonumber\\
&&\left.+\bar{\eta}_{{\bf{k}}+\frac{{\bf{q}}}{2},\sigma}\left(\nu_{n}\right)\bar{\eta}_{{\bf{k}}-\frac{{\bf{q}}}{2},\sigma}\left(\nu_{n}\right)\Delta^{\rm sc}_{\sigma}\right),
\label{Equation_44}
\end{eqnarray} 
where the superconducting order parameters are defined in the reals space as
\begin{eqnarray}
\bar{\Delta}^{\rm sc}_{\sigma}=2g\lim_{\tau'\rightarrow \tau}\left\langle \bar{\eta}_{\sigma}\left({\bf{r}}\tau\right)\bar{\eta}_{\sigma}\left({\bf{r}}\tau'\right)\right\rangle,
\nonumber\\
{\Delta}^{\rm sc}_{\sigma}=2g\lim_{\tau'\rightarrow \tau}\left\langle {\eta}_{\sigma}\left({\bf{r}}\tau'\right){\eta}_{\sigma}\left({\bf{r}}\tau\right)\right\rangle.
\label{Equation_45}
\end{eqnarray}
After those transformations, the total fermionic action of the system in Eq.(\ref{Equation_26}) becomes
\begin{eqnarray}
{\cal{S}}_{\rm gr}={\cal{S}}_{\rm t-B}+{\cal{S}}_{\rm U}+{\cal{S}}^{\rm el}_{\rm B}+{\cal{S}}_{\rm sc}.
\label{Equation_46}
\end{eqnarray}
It could be rewritten in the Nambu form as
\begin{eqnarray}
&&{\cal{S}}_{\rm gr}\left[\bar{\Psi},\Psi\right]=
\nonumber\\
&&=\frac{1}{2\beta{N}N_{\rm ph}}\sum_{\substack{{\bf{k}}{\bf{q}} \\ \nu_{n}\sigma}}\bar{\Psi}_{{\bf{k}}{\bf{q}},\sigma}\left(\nu_{n}\right){\cal{G}}^{-1}_{{\bf{k}}{\bf{q}},\sigma}\left(\nu_{n}\right){\Psi}_{{\bf{k}}{\bf{q}},\sigma}\left(\nu_{n}\right),
\nonumber\\
\label{Equation_47}
\end{eqnarray}
where we have introduced Nambu spinors $\bar{\Psi}_{{\bf{k}}{\bf{q}},\sigma}\left(\nu_{n}\right)$ and their complex conjugates $\left(\bar{\Psi}_{{\bf{k}}{\bf{q}},\sigma}\left(\nu_{n}\right)\right)^{T}$ as 
\begin{eqnarray}
{\Psi}_{{\bf{k}}{\bf{q}},\sigma}\left(\nu_{n}\right)&=\left(
\begin{array}{cccc}
\bar{a}_{{\bf{k}}+\frac{{\bf{q}}}{2},\sigma}\left(\nu_{n}\right)\\ \\
{a}_{{\bf{k}}-\frac{{\bf{q}}}{2},\sigma}\left(\nu_{n}\right)\\ \\
\bar{b}_{{\bf{k}}+\frac{{\bf{q}}}{2},\sigma}\left(\nu_{n}\right)\\ \\
{b}_{{\bf{k}}-\frac{{\bf{q}}}{2},\sigma}\left(\nu_{n}\right)
\end{array}\right).
\label{Equation_48}
\end{eqnarray}
The inverse Green's function matrix ${\cal{G}}^{-1}_{{\bf{k}}{\bf{q}},\sigma}\left(\nu_{n}\right)$, in Eq.(\ref{Equation_47}), is given in the following form
\begin{widetext}
\begin{eqnarray}
{\cal{G}}^{-1}_{{\bf{k}}{\bf{q}},\sigma}\left(\nu_{n}\right)=\left(
\begin{matrix}
  -i\nu_{n}-\mu_{1\rm eff} & -\Delta^{\rm sc}_{\sigma}  & -\gamma\left({\bf{k}}+\frac{{\bf{q}}}{2}\right) & 0\\
	-\bar{\Delta}^{\rm sc}_{\sigma} & i\nu_{n}+\mu_{1\rm eff}  & 0 & \gamma^{\ast}\left({\bf{k}}-\frac{{\bf{q}}}{2}\right)\\
	-\gamma^{\ast}\left({\bf{k}}+\frac{{\bf{q}}}{2}\right) & 0  & -i\nu_{n}-\mu_{2\rm eff} & -\Delta^{\rm sc}_{\sigma}\\
	0 & \gamma\left({\bf{k}}-\frac{{\bf{q}}}{2}\right)  & -\bar{\Delta}^{\rm sc}_{\sigma} & i\nu_{n}+\mu_{2\rm eff}\\
  \end{matrix}\right).
\label{Equation_49}
\end{eqnarray}
\end{widetext}
The effective chemical potentials $\mu_{i\rm eff}$ (with $i=1,2$) are defined as 
\begin{eqnarray}
\mu_{1\rm eff}=\mu-\frac{V}{2}-\frac{U}{2}\left\langle{n}_{a}\right\rangle,
\nonumber\\
\mu_{2\rm eff}=\mu+\frac{V}{2}-\frac{U}{2}\left\langle {n}_{b} \right\rangle.
\label{Equation_50}
\end{eqnarray}
Furthermore, $\gamma\left({\bf{k}}\right)$, in Eq.(\ref{Equation_49}), is the normalized dispersion relation in graphene 
\begin{eqnarray}
&&\gamma\left({\bf{k}}\right)=\gamma_{0}\sum_{\bm{\mathit{\delta}}}e^{-i{{\bf{k}}\bm{\mathit{\delta}}}}=
\nonumber\\
&&=\gamma_{0}\left(e^{-ik_{x}a_{0}}+2e^{i\frac{k_{x}a_{0}}{2}}\cos\left(\frac{\sqrt{3}}{2}k_{y}a_{0}\right)\right),
\label{Equation_51}
\end{eqnarray}
where $a_{0}$ is the lattice constant in graphene ($a_{0}=0.142$ nm). The external gate-potential $V$ induces the average charge density disequilibrium $\delta{\bar{n}}$ between the graphene sublattice sites positions $a$ and $b$
\begin{eqnarray}
\delta{\bar{n}}=\left\langle{n}_{a}\right\rangle-\left\langle{n}_{b}\right\rangle.
\label{Equation_52}
\end{eqnarray}
Moreover, we introduce the inverse filling coefficient $\kappa=1/n_{\rm fill}$ (where $n_{\rm fill}$ is the sum of average number of particles at the lattice sites $a$ and $b$) 
\begin{eqnarray}
\frac{1}{\kappa}=\left\langle{n}_{a}\right\rangle+\left\langle{n}_{b}\right\rangle=n_{\rm fill}.
\label{Equation_53}
\end{eqnarray}
Particularly, the value $\kappa=0.5$ corresponds to the half-filling limit, i.e., when there is only one electron per lattice site). Combining the expressions in Eqs.(\ref{Equation_52}) and (\ref{Equation_53}), the mean-field averages $\left\langle{n}_{a}\right\rangle$ and $\left\langle {n}_{b} \right\rangle$, figuring in the definitions of the effective chemical potentials in Eq.(\ref{Equation_49}), could be extracted as the functions of $\delta{\bar{n}}$ and $\kappa$. We obtain
\begin{eqnarray}
\left\langle{n}_{a}\right\rangle=\frac{1}{2}\left(\frac{1}{\kappa}+\delta{\bar{n}}\right),
\nonumber\\
\left\langle{n}_{b}\right\rangle=\frac{1}{2}\left(\frac{1}{\kappa}-\delta{\bar{n}}\right).
\label{Equation_54}
\end{eqnarray}
Next, we define the total superconducting order parameter in the system like the sum of the gap functions $\Delta^{\rm sc}_{\rm a}$ and $\Delta^{\rm sc}_{\rm b}$, i.e., $\Delta^{\rm sc}=\Delta^{\rm sc}_{\rm a}+\Delta^{\rm sc}_{\rm b}$, and we get the system of self-consistent equations for the chemical potential, average charge density difference and the superconducting order parameter $\Delta^{\rm sc}$. We have
\begin{eqnarray}
\frac{2}{NN_{\rm ph}}\sum_{{\bf{k}}{\bf{q}}}\sum^{4}_{i=1}\alpha_{i{\bf{k}}{\bf{q}}}n_{\rm F}\left(\mu-\varepsilon_{i{\bf{k}}{\bf{q}}}\right)=\frac{1}{\kappa},
\nonumber\\
\frac{2}{NN_{\rm ph}}\sum_{{\bf{k}}{\bf{q}}}\sum^{4}_{i=1}\beta_{i{\bf{k}}{\bf{q}}}n_{\rm F}\left(\mu-\varepsilon_{i{\bf{k}}{\bf{q}}}\right)=\delta{\bar{n}},
\nonumber\\
\frac{8g}{NN_{\rm ph}}\sum_{{\bf{k}}{\bf{q}}}\sum^{4}_{i=1}\gamma_{i{\bf{k}}{\bf{q}}}n_{\rm F}\left(\mu-\varepsilon_{i{\bf{k}}{\bf{q}}}\right)=-\Delta^{\rm sc},
\label{Equation_55}
\end{eqnarray}
where the ${\bf{k}}$-dependent coefficients $\alpha_{i{\bf{k}}}$, $\beta_{i{\bf{k}}}$ and $\gamma_{i{\bf{k}}}$ have been calculated in Appendix \ref{sec:Section_6}. The function $n_{\rm F}\left(x\right)$ in Eq.(\ref{Equation_55}) is Fermi-Dirac distribution function
\begin{eqnarray}
n_{\rm F}\left(x\right)=1/\left(e^{\beta\left(x-\mu\right)}+1\right),
\label{Equation_56}
\end{eqnarray}
and the energies $\varepsilon_{i{\bf{k}}{\bf{q}}}$ (with $i=1,...,4$) in the arguments of Fermi-Dirac function in Eq.(\ref{Equation_55}) define the electronic band structure of graphene. They have been obtained as
\begin{eqnarray}
\varepsilon_{i{\bf{k}}{\bf{q}}}=0.5\left(\mu_{1\rm eff}+\mu_{2\rm eff}+\left(-1\right)^{i}\sqrt{E_{1{\bf{k}}{\bf{q}}}-2\sqrt{E_{2{\bf{k}}{\bf{q}}}}}\right),
\nonumber\\
\label{Equation_57}
\end{eqnarray}
for $i=1,2$, and 
\begin{eqnarray}
\varepsilon_{j{\bf{k}}{\bf{q}}}=0.5\left(\mu_{1\rm eff}+\mu_{2\rm eff}+\left(-1\right)^{j}\sqrt{E_{1{\bf{k}}{\bf{q}}}+2\sqrt{E_{2{\bf{k}}{\bf{q}}}}}\right),
\nonumber\\
\label{Equation_58}
\end{eqnarray}
for $j=3,4$. The functions $E_{1{\bf{k}}{\bf{q}}}$ and $E_{2{\bf{k}}{\bf{q}}}$, in Eqs.(\ref{Equation_57}) and (\ref{Equation_58}) are
\begin{eqnarray}
E_{1{\bf{k}}{\bf{q}}}&=&2\left|\gamma\left({\bf{k}}+\frac{{\bf{q}}}{2}\right)\right|^{2}+2\left|\gamma\left({\bf{k}}-\frac{{\bf{q}}}{2}\right)\right|^{2}
\nonumber\\
&&+\left(\mu_{1\rm eff}-\mu_{2\rm eff}\right)^{2}-4\left|\Delta^{\rm sc}\right|^{2},
\label{Equation_59}
\end{eqnarray}
and
\begin{eqnarray}
E_{2{\bf{k}}{\bf{q}}}&=&\left(\left|\gamma\left({\bf{k}}+\frac{{\bf{q}}}{2}\right)\right|^{2}-\left|\gamma\left({\bf{k}}-\frac{{\bf{q}}}{2}\right)\right|^{2}\right)^{2}
\nonumber\\
&&-4\left|\Delta^{\rm sc}\right|^{2}\left(\gamma\left({\bf{k}}+\frac{{\bf{q}}}{2}\right)\gamma^{\ast}\left({\bf{k}}-\frac{{\bf{q}}}{2}\right)\right.
\nonumber\\
&&\left.+\gamma^{\ast}\left({\bf{k}}+\frac{{\bf{q}}}{2}\right)\gamma\left({\bf{k}}-\frac{{\bf{q}}}{2}\right)+\left(\mu_{1\rm eff}-\mu_{2\rm eff}\right)^{2}\right).
\nonumber\\
\label{Equation_60}
\end{eqnarray}
In the next section, we will show how the band-gap is opening in the electronic structure of graphene, even at $V=0$, due to superconducting pairing in the system.  
%
\section{\label{sec:Section_4} Numerical results and discussion}
\begin{figure}
	\includegraphics[scale=0.4]{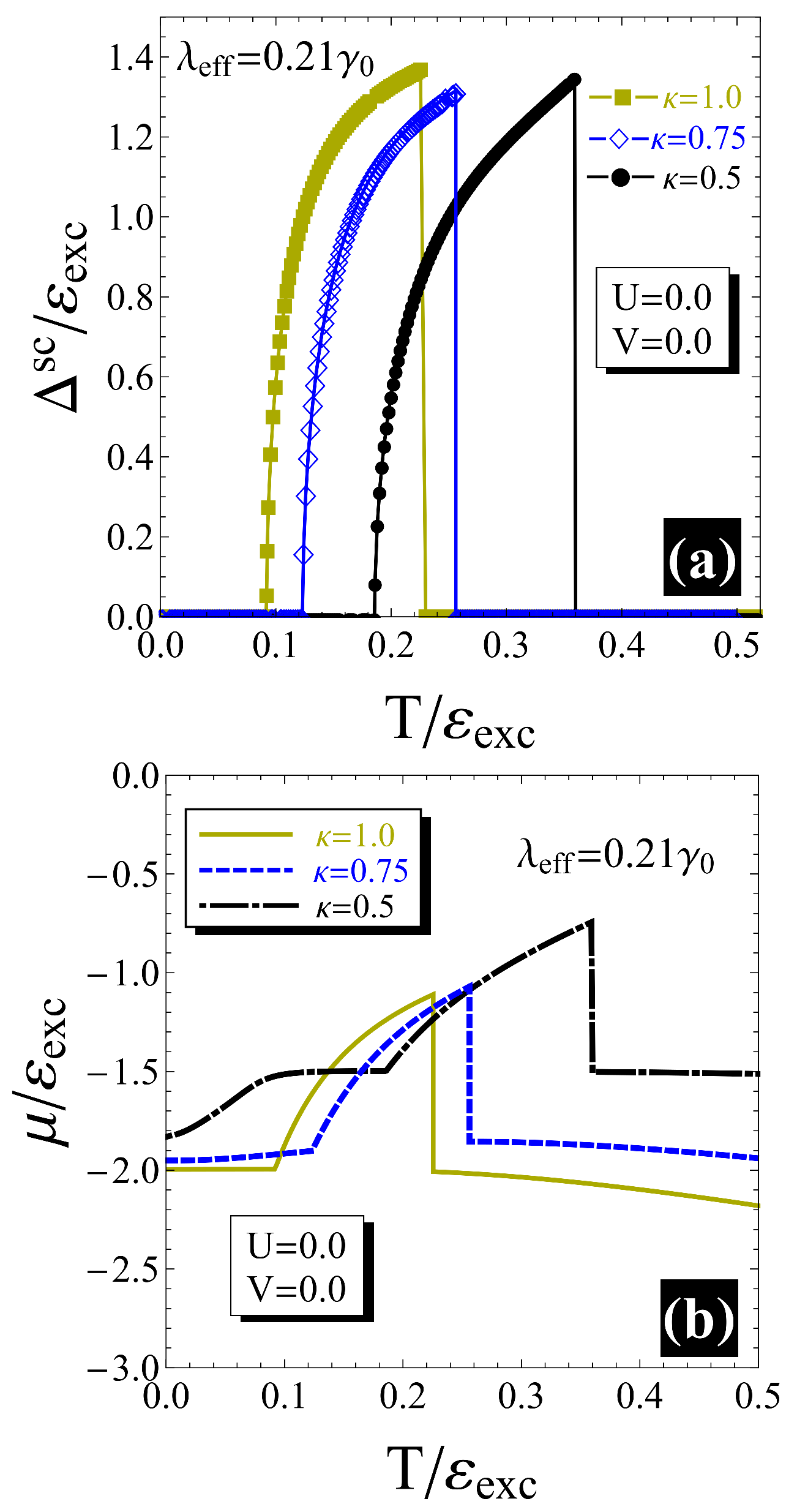}
	\caption{\label{fig:Fig_4} The superconducting order parameter $\Delta^{\rm sc}$ (see in panel (a)), the chemical potential $\mu$ (see in panel (b)), calculated from Eq.(\ref{Equation_55}), as a function of temperature. The effective electron-phonon interaction parameter has been fixed at $\lambda_{\rm eff}=0.21\gamma_0=0.63$ eV and different values of the inverse filling coefficient $\kappa$ have been considered in the calculations.
	}
\end{figure}

%
The system of equations, in Eq.(\ref{Equation_55}), has been solved exactly at the vicinity of the point ${\bf{k}}={\bf{K}}$ and by considering the ${\bf{q}}$-excitation vector in the interval ${\bf{q}}\in\left[0.9{\bf{K}},1.1{\bf{K}}\right]$. The finite difference approximation has been employed to this aim \cite{cite_41}, that retains Newton's fast convergence algorithm.
In Fig.~\ref{fig:Fig_4}, we have been calculated the superconducting order parameter $\Delta^{\rm sc}$ (see in panel (a))and (see in panel (b)), as a function of temperature, for different values of the inverse filling coefficient $\kappa$. The effective electron-phonon interaction coefficient has been fixed at $\lambda_{\rm eff}=0.21\gamma_0=0.63$ eV. 
%
%
%
%
The unbiased graphene has been considered, and the average charge density difference $\delta{\bar{n}}$ (see the second equation, in Eq.(\ref{Equation_55})) vanishes in this case $\delta{\bar{n}}=0$, because $V=0$. We have found a sufficiently large critical region $\Delta{T}=T_{\rm C2}-T_{\rm C1}$ on the temperature axis, where the superconducting gap parameter is not zero. The highest transition temperatures, into the superconducting state, were obtained in the case of half-filling with $\kappa=0.5$ (see the black curves in panels (a) and (b), in Fig.~\ref{fig:Fig_4}). In this case, the first transition temperature $T_{\rm C1}$, at which the superconducting gap appears was found surprisingly high, of order $T_{\rm C1}=0.185\varepsilon_{\rm exc}=121.7$ K, and the second transition temperature $T_{C2}$ at which the gap disappears, and the system passes to the normal states, is $T_{C2}=0.359\varepsilon_{\rm exc}=236.2$ K (see the region, in panel (b), where the gap jumps drastically to zero). The maximum value, of the superconducting gap, was obtained at the edge $T=T_{\rm C2}$ of transition into the normal state, and we get $\Delta^{\rm sc}_{\rm max}=1.351\varepsilon_{\rm exc}=76.6$ meV.   
%
%
\begin{figure}
	\includegraphics[scale=0.37]{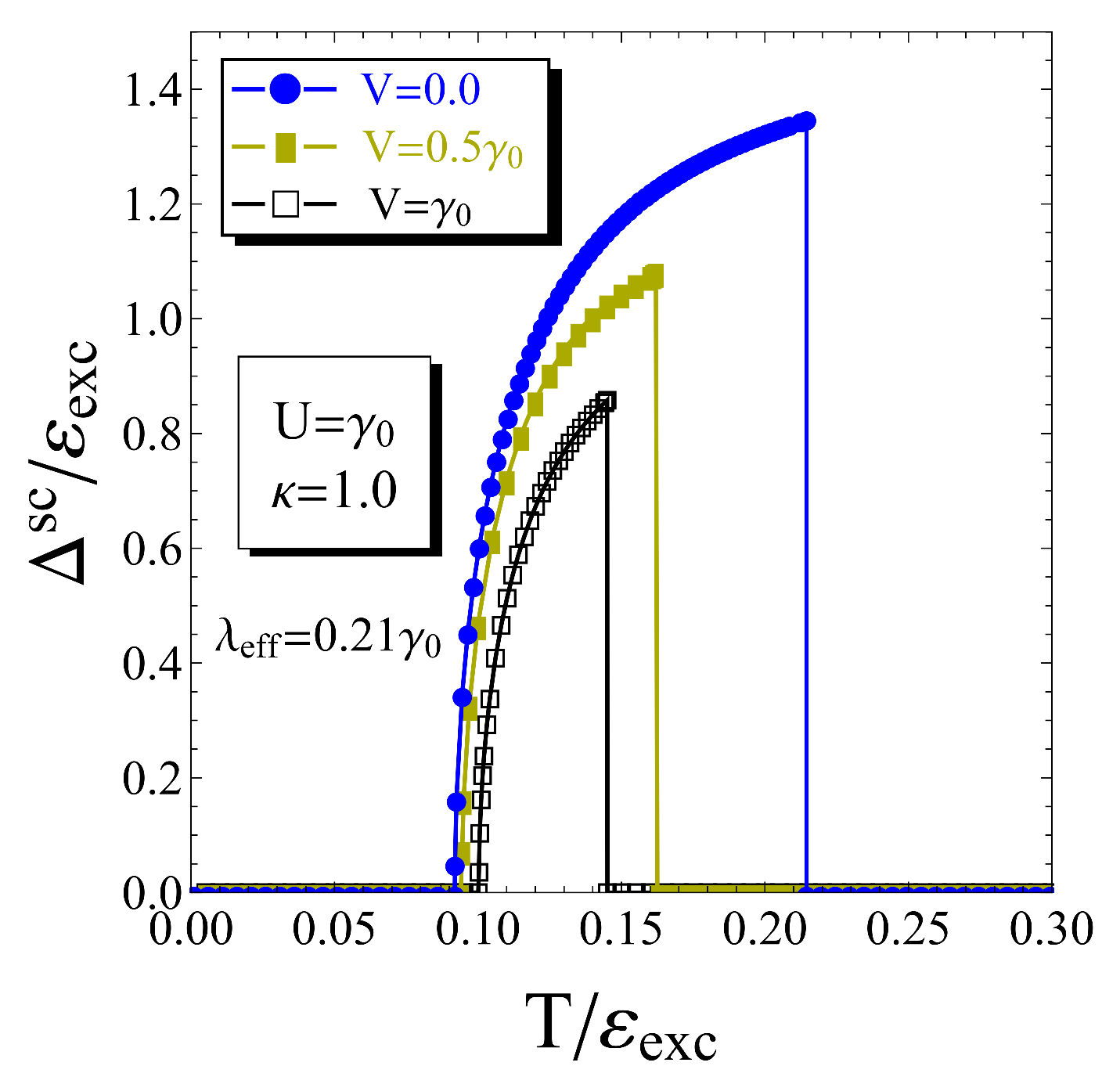}
	\caption{\label{fig:Fig_5}(Color online) The superconducting order parameter $\Delta^{\rm sc}$, calculated from 
     Eq.(\ref{Equation_55}), as a function of temperature. The effective electron-phonon interaction parameter has been set at the value $\lambda_{\rm eff}=0.21\gamma_0=0.63$ eV, and different values of the external gate voltage $V$ have been considered in the calculations.
	}
\end{figure} 
%
%
Those critical temperatures decrease when increasing the coefficient $\kappa$, i.e., when passing to the limit of partial-filling (when the sum of average number of particles at lattice sites $a$ and $b$ has to be fractional). For example, at $\kappa=0.75$, we find $T_{\rm C1}=0.123\varepsilon_{\rm exc}=80.93$ K, $T_{\rm C2}=0.214\varepsilon_{\rm exc}=168.5$ K, and $\Delta^{\rm sc}_{\rm max}=74.58$ meV. Next, for $\kappa=1.0$, we find $T_{\rm C1}=0.088\varepsilon_{\rm exc}=57.96$ K and $T_{\rm C2}=0.225\varepsilon_{\rm exc}=148.37$ K and $\Delta^{\rm sc}_{\rm max}=77.52$ K. The enhancement of the superconducting phase transition is strongly related to the solution of the chemical potential in graphene. In the transition region (see in panel (b), in Fig.~\ref{fig:Fig_4}), the superconducting phase $T\in\left[T_{\rm C1},T_{\rm C2}\right]$ starts exactly when $\left|\mu\right|$ experiences the unprecedented decrease. We see that $\left|\mu\right|$ in the transition region gets strongly decreased (because $\mu<0$), and, as a consequence, the grand thermodynamic potential $\Phi={\cal{F}}-\mu{N}$ gets suddenly minimized in that case, which is sign of the passage to a new state of matter, i.e., the superconducting pairing state, in our case. Moreover, the small values of $\left|\mu\right|$ favor the single-particle excitations above or below Fermi sea (creation or annihilation) and consequently foster the possibilities of emergence of the superconducting state, the latest is being a natural consequence of a complex of complicated many-body interactions in the system which need an appropriate number of single-particle excitations in the system to create the ordered phase. 

The appearance of superconductivity, here, has strong relation with the considered ${\bf{q}}$-excitations in our model. We have quantized the single-particle excitation into two discrete energies near Fermi level, which could be explained in the sense similar to that of the discretization of energy levels in reduced dimensions (known as quantum size effect (QSE) \cite{cite_42}). It is known that the oscillations of the average electron density near Fermi level are direct consequences of the QSE \cite{cite_43}.

In Fig.~\ref{fig:Fig_5}, we calculated superconducting order parameter $\Delta^{\rm sc}$, for various values of the external gate-potential $V$. The partial-filling case was considered, with $\kappa=1.0$. The on-site Hubbard interaction potential $U$ has been fixed at $U=\gamma_0=3$ eV. We see that the superconducting transition regions $\left[T_{\rm C1},T_{\rm C2}\right]$ become narrow when increasing the gate-potential $V$. In the matter of fact, that means that the external gate-potential has a destructive effect on the SC phase and order parameter $\Delta^{\rm sc}$.

\begin{figure}
	\includegraphics[scale=0.42]{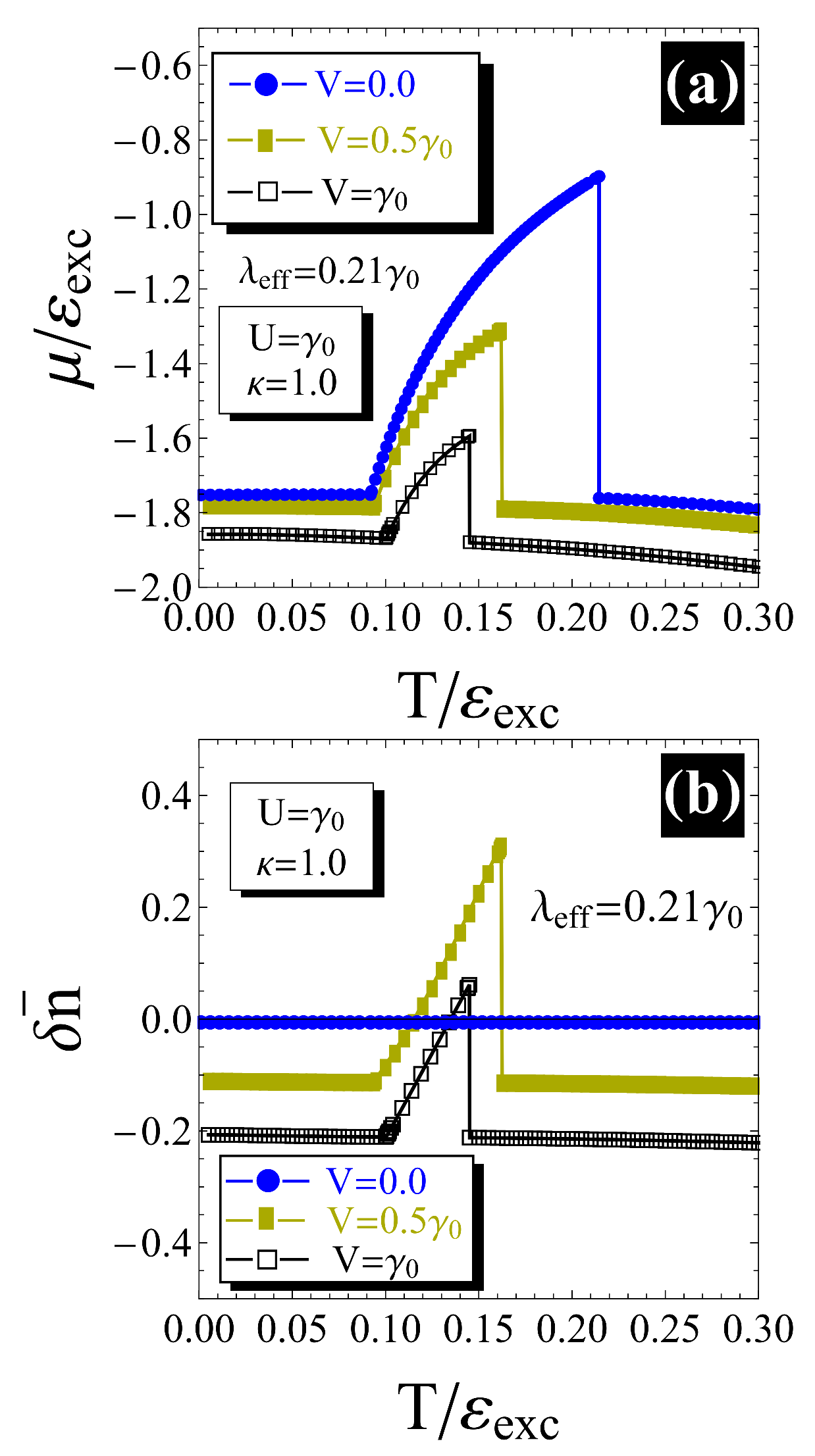}
	\caption{\label{fig:Fig_6}(Color online) The chemical potential $\mu$ (panel (a)), and the average charge density imbalance $\delta{n}$ (panel (b)), calculated from Eq.(\ref{Equation_55}), as a function of temperature. The effective electron-phonon interaction parameter has been fixed at $\lambda_{\rm eff}=0.21\gamma_0=0.63$ eV, and different values of the external gate voltage $V$ have been considered. All energy parameters are normalized to the single-particle ${\bf{q}}$-excitation energy $\varepsilon_{\rm exc}$.
	}
\end{figure}

In Fig.~\ref{fig:Fig_6}, we have shown the temperature dependence of the chemical potential (see in panel (a)) and average charge density imbalance (see in panel (b)), for the same values of the other input parameters as in Fig.~\ref{fig:Fig_5}. We observe in panel (b), in Fig.~\ref{fig:Fig_6}, that for $V\neq 0$ and beyond the SC transition region, $\delta{\bar{n}}$ is nearly a constant and negative function of temperature. That means that there exists an imbalance between the average electron concentrations on sites $a$ and $b$ satisfying the inequality $\bar{n}_{a}<\bar{n}_{b}$. In the transition region $\left[T_{\rm C1},T_{\rm C2}\right]$, the function $\delta{\bar{n}}$ increases with $T$, passing through the charge neutrality point when $\delta{\bar{n}}=0$, then, at $T=T_{\rm C2}$, it jumps to its solution line for $T<T_{\rm C1}$. In the SC transition region, the function $\delta{\bar{n}}$ acquires also large positive values which means the inequality $\bar{n}_{a}>\bar{n}_{b}$ takes place (see plot in dark yellow, corresponding to the case $V=0.5\gamma_0=1.5$ eV).  

\begin{figure}
	\includegraphics[scale=0.38]{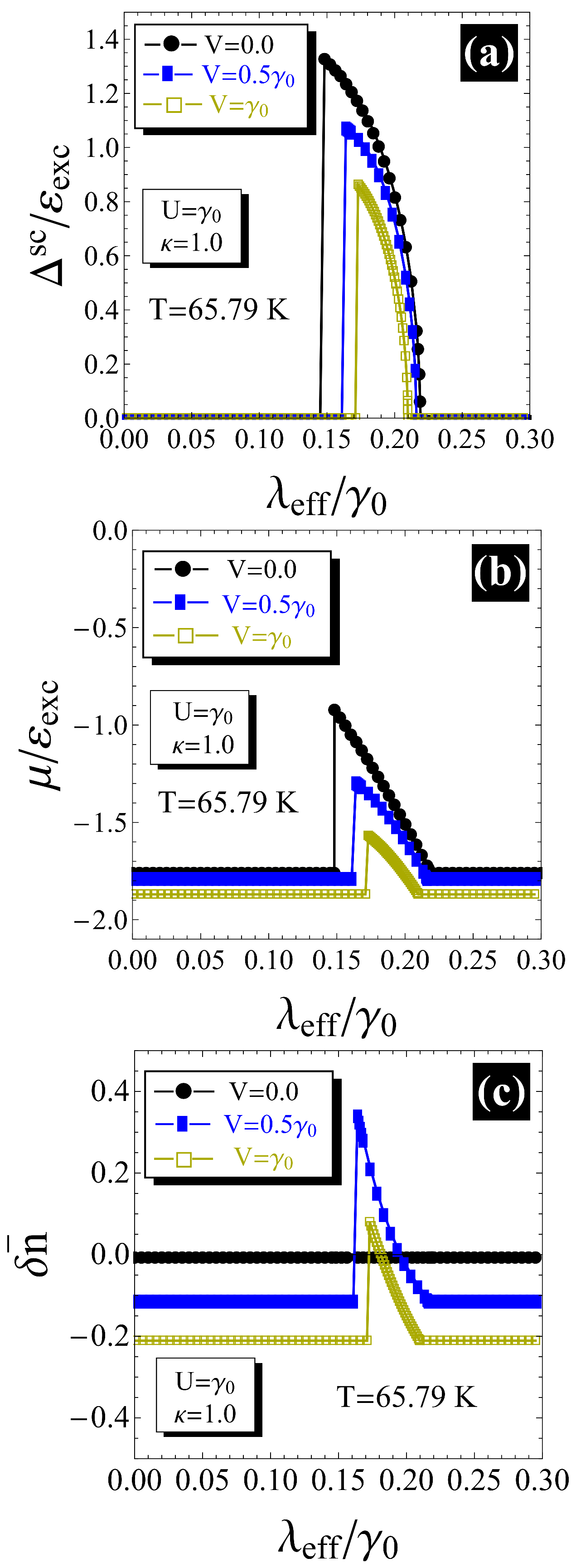}
	\caption{\label{fig:Fig_7}(Color online) The superconducting order parameter $\Delta^{\rm sc}$ (panel (a)), chemical potential $\mu$ (panel (b)) and average density difference $\delta{\bar{n}}$ (panel (c)), calculated from Eq.(\ref{Equation_55}), as a function of the electron-phonon interaction parameter $\lambda_{\rm eff}$. The temperature has been set at the value $T=0.1\varepsilon_{\rm exc}=65.79$ K, and different values of the external gate voltage $V$ have been considered during calculations.}
\end{figure}

In Fig.~\ref{fig:Fig_7}, (see panels (a)-(c)), we have shown the dependence of the calculated quantities on the effective electron-phonon interaction parameter $\lambda_{\rm eff}$. The calculations have been performed at $T=0.1\varepsilon_{\rm exc}=65.79$ K. The results, in Fig.~\ref{fig:Fig_7}, have been obtained for the case of fractional filling with $\kappa=1.0$ (well beyond the half-filling limit $\kappa=0.5$). In fact, the value $\kappa=1.0$ could correspond the case when the lattice sites $a$ are occupied by the electrons and sites $b$ occupied by the holes only. The local Hubbard interaction $U$ was set at the value $U=\gamma_0=3$ eV. We have found a narrow critical region of the coupling parameter $\lambda_{\rm eff}$ ($\lambda_{\rm eff}\in\left[\lambda_{\rm C1},\lambda_{\rm C2}\right]$), where the gap $\Delta^{\rm sc}$ is non-zero, mainly, $\lambda_{\rm eff}\in\left[0.148\gamma_0,0.219\gamma_0\right]=\left[0.44,0.657\right]$ eV, for the case $V=0$ (see in panel (a)). Here, again, the role of the chemical potential is essential (see in panel (b)) and the obtained behavior of the superconducting order parameter (see in panel (a), in Fig.~\ref{fig:Fig_7}) reminiscent on the existence of quantum critical point in Ref.\onlinecite{cite_12}. We see in panel (b), in Fig.~\ref{fig:Fig_7}, that, in the region of superconducting pairing, the obtained values of $\left|\mu\right|$ become drastically diminished, which involves the rule for the single-particle excitations in the system and the consequent pairing interaction. The plots for the average particle density difference $\delta{\bar{n}}$ have been presented in panel (c), in Fig.~\ref{fig:Fig_7}. Here, the charge neutrality at $V\neq 0$ occurs only in the SC transition region $\lambda_{\rm eff}\in\left[\lambda_{\rm C1},\lambda_{\rm C2}\right]$.         
%
%
%
%
%
%
%
\begin{figure}
	\includegraphics[scale=0.328]{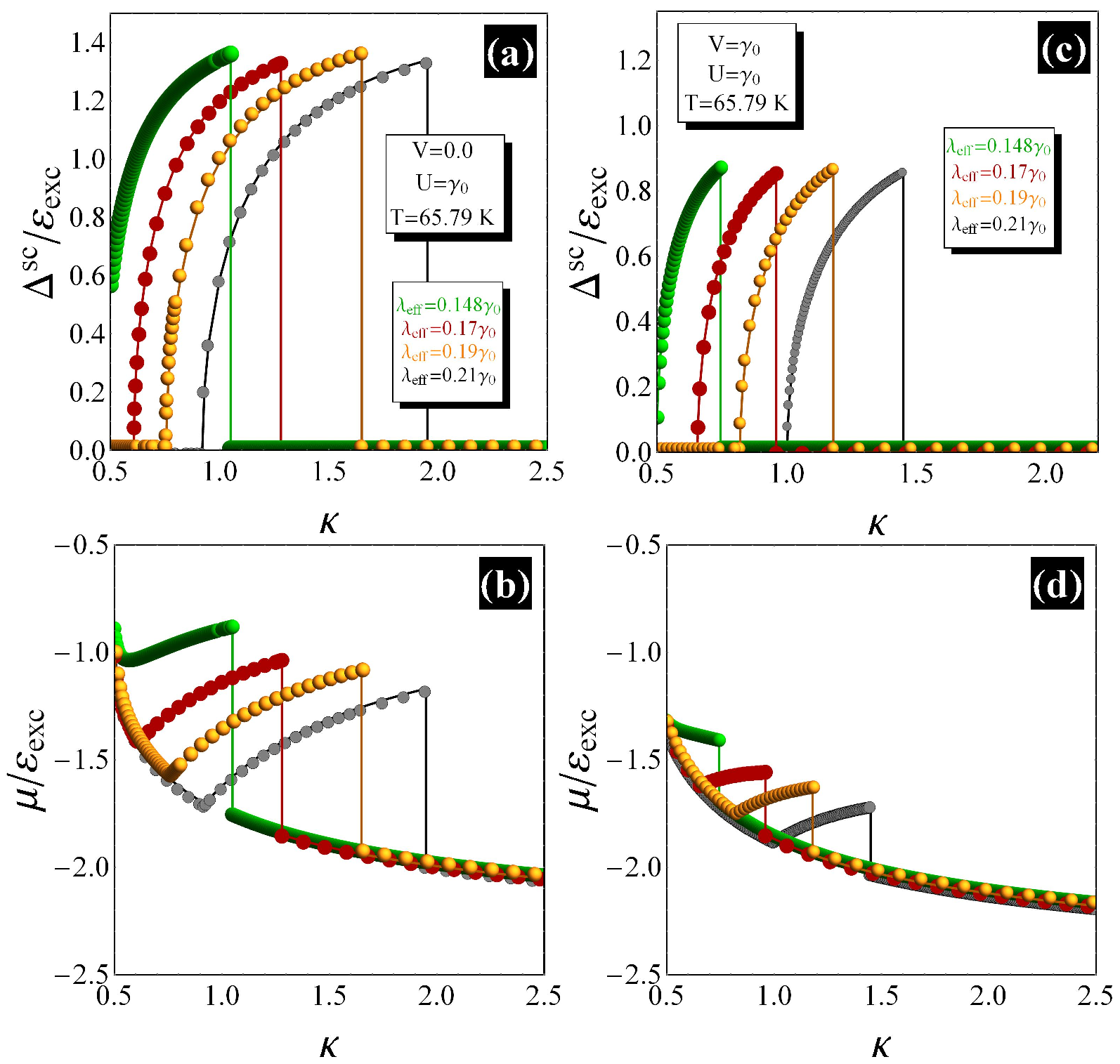}
	\caption{\label{fig:Fig_8}(Color online) The superconducting order parameter $\Delta^{\rm sc}$ (panels (a) and (c)), and the chemical potential in the system (panels (b) and (d)), calculated as a function of the inverse filling coefficient $\kappa=1/n_{\rm fill}$. Two different values of potential $V$ have been employed. Plots in panels (a) and (b) correspond to the case of the unbiased system with $V=0$, while the value $V=\gamma_0=3$ eV was employed in panels (c) and (d). Hubbard interaction potential has been fixed at $U=\gamma_0=3$ eV. Various limits of the electron-phonon interaction parameter $\lambda_{\rm eff}$ have been considered, and calculations have been performed at $T=0.1\varepsilon_{\rm exc}=65.79$ K.   
	}
\end{figure} 
%
%
%
\begin{figure}
	\includegraphics[scale=0.6]{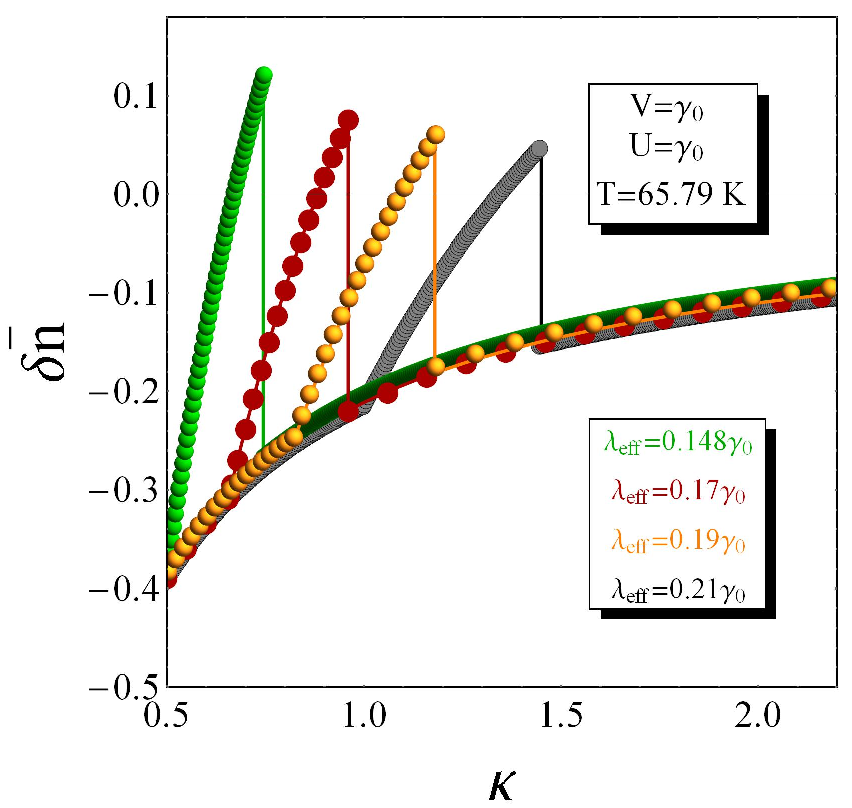}
	\caption{\label{fig:Fig_9}(Color online) The average charge density difference 
	$\delta{\bar{n}}=\bar{n}_{\rm a}-\bar{n}_{\rm b}$, calculated as a function of the inverse filling coefficient $\kappa=1/n_{\rm fill}$. Two different limits of the effective electron-phonon interaction parameter have been considered, for the case $V=\gamma_0=3$ eV. The Hubbard interaction potential has been fixed at $U=\gamma_0=3$ eV, and the calculations have been performed at $T=0.1\varepsilon_{\rm exc}=65.79$ K.
	}
\end{figure} 
%
%

For the experimental observation of the such enhanced values of the superconducting transition critical temperature, obtained in Figs.~\ref{fig:Fig_4}-\ref{fig:Fig_7}, the electron-phonon coupling constant should be carefully determined, in the appropriate ranges given in the figures. In turn, this could be realized via the experimental measurements of the phonon density of states, using nuclear resonant inelastic scattering (NRIXS) \cite{cite_44}. That technique enables determine the electron-phonon coupling constant $\lambda_{\rm eff}$ (which could be compared to the results in Fig.~\ref{fig:Fig_7}) and possible phonon-softening, which is, in general, responsible for the substantial enhancement of the superconducting transition critical temperature \cite{cite_45}. 

In Fig.~\ref{fig:Fig_8}, we have shown the calculations of the SC order parameter (see in panels (a) and (c)) and chemical potential (see in panels (b) and (d)) in the system as a function of the inverse filling coefficient $\kappa=1/n_{\rm fill}$. Remember, that minimal value for the parameter $\kappa$ is $\kappa=0.5$, which corresponds to the case of half-filling with $n_{\rm fill}=2.0$. The calculations have been performed at $T=65.79$ K and for two different values of the external gate-potential, mainly for $V=0$ (see in panels (a) and (b)) and for $V=\gamma_0=3$ eV (see in panels (c) and (d)). The on-site Hubbard potential has been fixed at $U=\gamma_0=3$ eV and different values of the electron-phonon interaction parameter $\lambda_{\rm eff}$ have been used, from the interval $\lambda_{\rm eff}\in\left[0.148\gamma_0,0.21\gamma_0\right]$, from weak to strong coupling limit. 
We see, in panels (a) and (c), that for each value of the interaction parameter $\lambda_{\rm eff}$, we have well separated domains of SC phases, which are displacing to the right on the $\kappa$-axis (toward small numbers of the inverse on-site occupation), when increasing the effective coupling parameter $\lambda_{\rm eff}$, from weak to strong coupling limit.

We observe, also, that for the unbiased case $V=0$ and at the strong electron-phonon coupling limit (see plots in black, in panels (a) and (c) with $\lambda_{\rm eff}=0.21\gamma_0=0.63$ eV) the superconducting phase could be formed with a large number of possible values of $\kappa$, and, there is no solution for $\Delta^{\rm sc}$ at the half-filling $\kappa=0.5$. Particularly, the coefficient $\kappa$ varies, in this case, within the interval $\kappa\in\left[0.915, 2.045\right]$ corresponding to the total filling coefficient $n_{\rm fill}\in\left[0.489,1.092\right]$. 
For all values of $n_{\rm fill}$, within this interval, the superconducting state exists (see the curves for $\Delta^{\rm sc}$, in panel (a), in Fig.~\ref{fig:Fig_8}). Meanwhile, in the weak electron-phonon interaction limit, with $\lambda_{\rm eff}=0.148\gamma_0=0.44$ eV (see the curves in green, in panels (a) and (c)), the region of the superconducting pairing phase becomes narrow and displaced to the left on the $\kappa$-axis. In this case, the values of $\kappa$, corresponding to the solutions of $\Delta^{\rm sc}$, are well located in the interval $\kappa\in\left[0.5,1.05\right]$ (or, the corresponding possible values of the electron filling coefficient $n_{\rm fill}$ are given in the interval $n_{\rm fill}\in\left[0.952,2.0\right]$), and there is a non-zero solution for $\Delta^{\rm sc}$ at half-filling $\kappa=0.5$, with $\Delta^{\rm sc}=0.562\varepsilon_{\rm exc}=31.9$ meV. For the biased case (see in panels (c) and (d)), the SC region is formed with a gradually reduced interval of $\kappa$. Namely, for $\lambda_{\rm eff}=0.21\gamma_0$, we have $\kappa\in\left[1.002,1.45\right]$ (or $n_{\rm fill}\in\left[0.689,0.998\right]$), while in the weak coupling limit, for $\lambda_{\rm eff}=0.148\gamma_0$, we get $\kappa\in\left[0.5,0.745\right]$ (corresponding to $n_{\rm fill}\in\left[1.342,2.0\right]$). We conclude that for small values of $\lambda_{\rm eff}$, the SC state could be realized with a larger interval of values $n_{\rm fill}$, which leads to the formation of the superconducting state, being almost a complex of a huge number of inter-particle correlations.   

In the case of the unbiased graphene, the values of $\left|\mu\right|$ are smaller (see in panel (b), in Fig.~\ref{fig:Fig_8}) than in the case of the biased sample (see in panel (d)). The smaller values of $\left|\mu\right|$ favor the formation of the SC state, at it was mentioned earlier. Therefore the magnitude of the corresponding SC order parameter $\Delta^{\rm sc}$ is larger in this case (see in panel (a)), compared to the biased case, shown in panel (c), in Fig.~\ref{fig:Fig_8}. At some critical values of the parameter $\kappa$ the chemical potential jumps into the solutions lines when $\Delta^{\rm sc}=0$. 
%
%
\begin{figure}
	\includegraphics[scale=0.15]{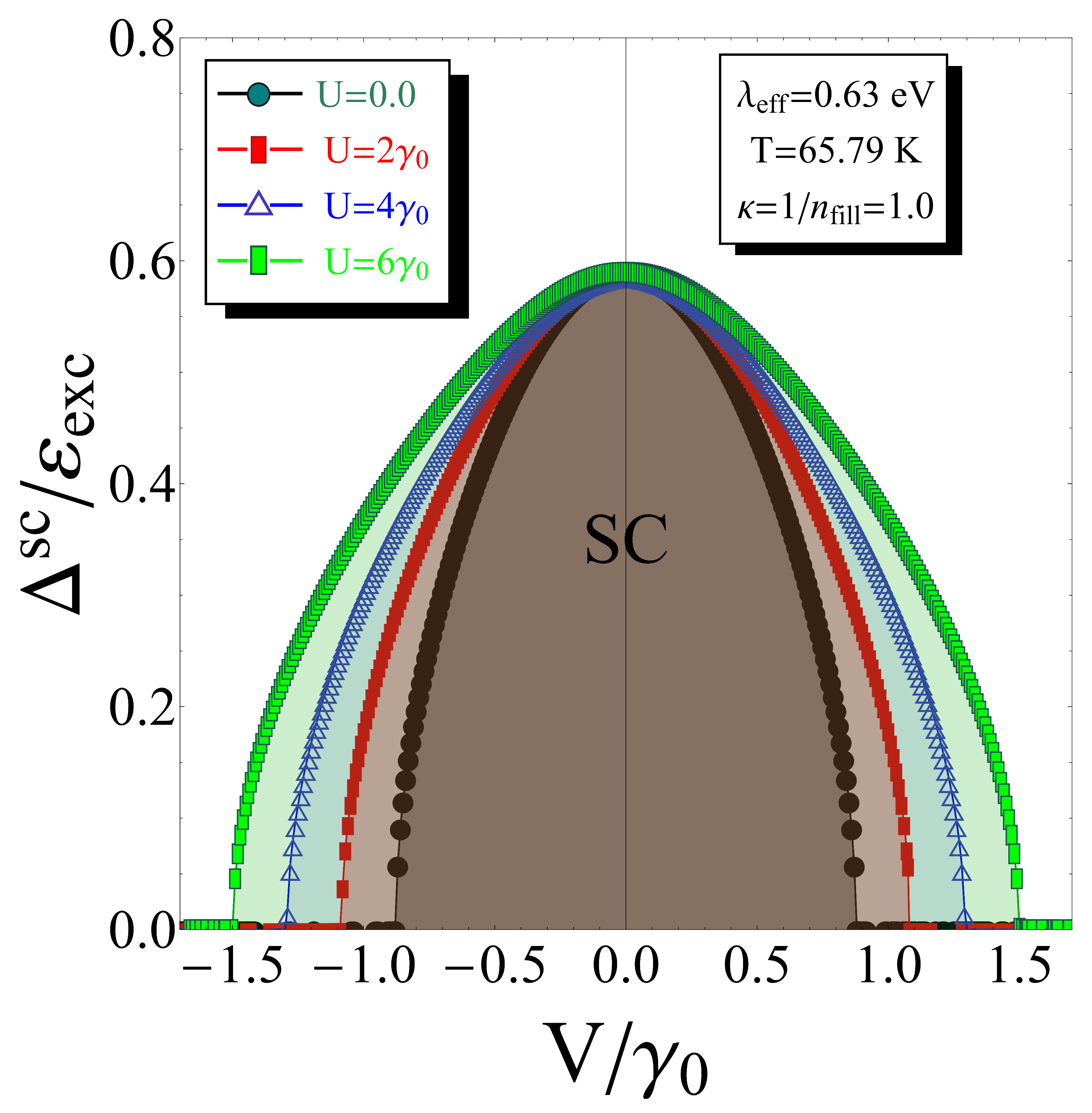}
	\caption{\label{fig:Fig_10}(Color online) The superconducting order parameter $\Delta^{\rm sc}$, calculated as a function the applied gate-potential $V$. Different values of the local Hubbard-$U$ potential have been considered. The inverse filling coefficient is set at the value $\kappa=1.0$ (partial-filling), and the strong electron-phonon interaction limit has been considered with $\lambda_{\rm eff}=0.21\gamma_0=0.63$ eV. The calculations have been performed at $T=0.1\varepsilon_{\rm exc}=65.79$ K.}
\end{figure} 
%
%

Furthermore, in Fig.~\ref{fig:Fig_9}, we present the calculated average charge density difference $\delta{\bar{n}}$ as a function of the inverse filling coefficient $\kappa$. The other parameters in the system are the same, as in Fig.~\ref{fig:Fig_8}. The triangular-like domains, in Fig.~\ref{fig:Fig_9}, correspond to the SC pairing in the system. We see that charge neutrality takes place, in the SC region, practically for all values of the electron-phonon interaction parameter $\lambda_{\rm eff}$. Thus, at given unique values of the inverse filling coefficient $\kappa=\kappa_{0}$,  the system returns into its charge equilibrium state $\delta{n}\left(\kappa_0\right)=0$. 

\begin{figure}
	\includegraphics[scale=0.35]{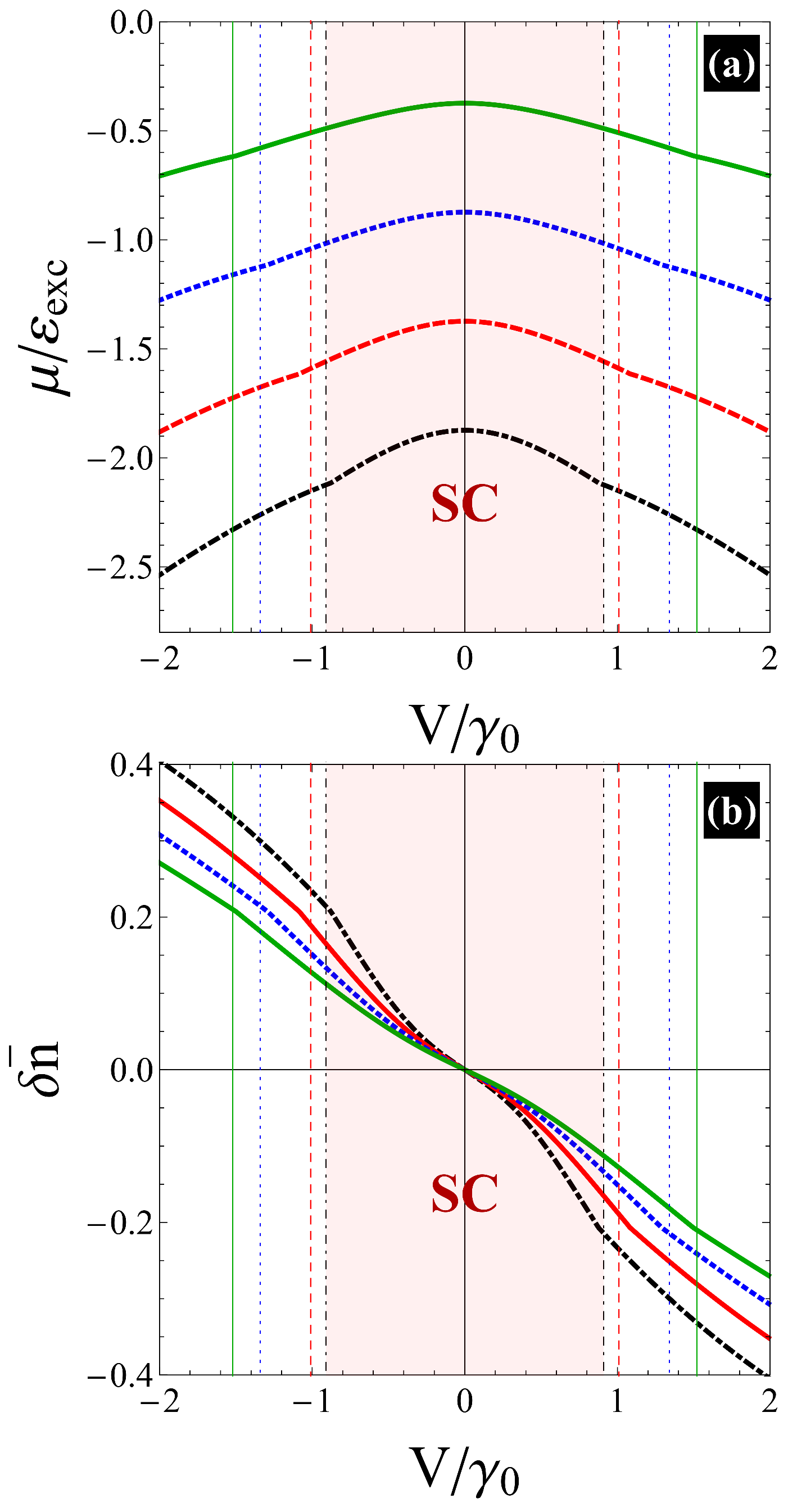}
	\caption{\label{fig:Fig_11}(Color online) The chemical potential (see in panel (a)) and average charge density imbalance (see panel (b)), as a function of applied gate-potential $V$. Different values of the local Hubbard-$U$ potential have been considered. The inverse filling coefficient is set at the value $\kappa=1.0$ (the case of partial-filling), and the strong electron-phonon interaction limit has been considered with $\lambda_{\rm eff}=0.21\gamma_0=0.63$ eV. The calculations have been done at $T=0.1\varepsilon_{\rm exc}=65.79$ K. The vertical lines separate the normal and superconducting states (SC) in the system, for each value of Hubbard-$U$ potential. The light-red domains indicate the SC phase corresponding to the value $U=0$ eV.}
\end{figure}

Interesting results have been obtained in Figs.~\ref{fig:Fig_10} and ~\ref{fig:Fig_11}, where we have calculated the $\Delta^{\rm sc}$ (Fig.~\ref{fig:Fig_10}), $\mu$ and $\delta{\bar{n}}$ (Fig.~\ref{fig:Fig_11}), as a function of the applied gate-potential $V$. Different limits of the local Hubbard-$U$ interaction have been considered, from the interval $U\in\left[0,6\gamma_0\right]=\left[0,18\right]$ eV. The calculations have been performed at the partial-filling case with $\kappa=1.0$, and temperature was fixed at $T=0.1\varepsilon_{\rm exc}=65.79$ K. We see, in Fig.~\ref{fig:Fig_10}, that the triplet pairing order parameter $\Delta^{\rm sc}$ decreases when augmenting the external gate-potential $\left|V\right|$. Meanwhile, for the large values of the local Coulomb potential $U$, the suppression of the SC state happens at the higher values $\left|V_{\rm C}\right|$ of the applied gate-potential. Thus, the large values of the interaction potential $U$ enrich the superconducting pairing, and the superconducting domain becomes larger (see superconducting phases in black and green, corresponding to the limits $U=0$ eV and $U=6\gamma_0=18$ eV). For example at $U=0$ (see the curve in black, in Fig.~\ref{fig:Fig_10}) the superconducting order parameter vanishes at $V_{\rm C}=0.91\gamma_0=2.73$ eV, while for the large-$U$ limit, with $U=18$ eV, we have $\Delta^{\rm sc}=0$ at $V_{\rm C}=1.52\gamma_0=4.56$ eV. Large values of $U$ correspond to the strong repulsion between the electrons with spins $\uparrow$ and $\downarrow$, which increases the possibility of the spin-triplet SC pairing states in the system. The solutions for the chemical potential and average charge density difference have been shown in panels (a) and (b), in Fig.~\ref{fig:Fig_11}. Here, again, the small values of $\left|\mu\right|$, obtained for the case of large $U$, minimize the grand canonical potential of the system and, consequently, leads to an increase of the possible spin-triplet ${\bf{q}}$-pairing states in the system. The vertical lines shown in the pictures, in Fig.~\ref{fig:Fig_11}, indicate the critical values of the gate-potential $V_{\rm C}$ (for each $U$) at which the system passes to the normal state phase. The domains in light-red, in Fig.~\ref{fig:Fig_11}, indicate purely SC state, corresponding to the limit $U=0$ eV. We see that, in the case $U=6\gamma_0=18$ eV, the normal state is attained at the larger value of the applied gate-potential $V$. The variation of the charge-imbalance $\delta{\bar{n}}$ is shown in panel (b), in Fig.~\ref{fig:Fig_11}. We observe that the calculated values of the function $\delta{\bar{n}}$ change the sign with $V$, and the charge neutrality occurs only in the case $V=0$ eV. Another interesting result, here, is that the average charge fluctuations $\delta{\bar{n}}$ as a function of $V$ are small in the case of strong $U$ interaction potential (see line in green on the picture, in panel (b)). That corresponds to the situation when the electrons get strongly localized on their positions in space. Therefore, large-$U$ limit corresponds to the strong localization of the electrons in the lattice \cite{cite_46, cite_47, cite_48, cite_49}. As a consequence, the reduction of the average charge density fluctuations enhances the SC order in the system, as we realize in Fig.~\ref{fig:Fig_10} (see plot in green, in Fig.~\ref{fig:Fig_10}) and the ${\bf{q}}$-superconductivity is more pronounced in that case. The corresponding behavior of the chemical potential is presented in panel (a), in Fig.~\ref{fig:Fig_11}. We observe, also, in panel (b), in Fig.~\ref{fig:Fig_11}, that the average charge fluctuations $\delta{\bar{n}}$, from charge neutrality value $\delta{\bar{n}}=0$, are larger for the high values of the applied gate-potential $V$ (for all given $U$). Therefore, those large fluctuations are responsible for the destabilization and consequent destruction of the triplet ${\bf{q}}$-superconducting state in graphene. We see that the curves $\delta{\bar{n}}$ get slightly inflamed at the critical values $V_{\rm C}$ of the gate-potential $V$ (see vertical lines, in panels (a) and (b), in Fig.~\ref{fig:Fig_11}). Above calculated values $V_{\rm C}$, in Figs.~\ref{fig:Fig_10} and ~\ref{fig:Fig_11}, the system passes into the normal state.            
%
%
%
%
\begin{figure}
\includegraphics[scale=0.225]{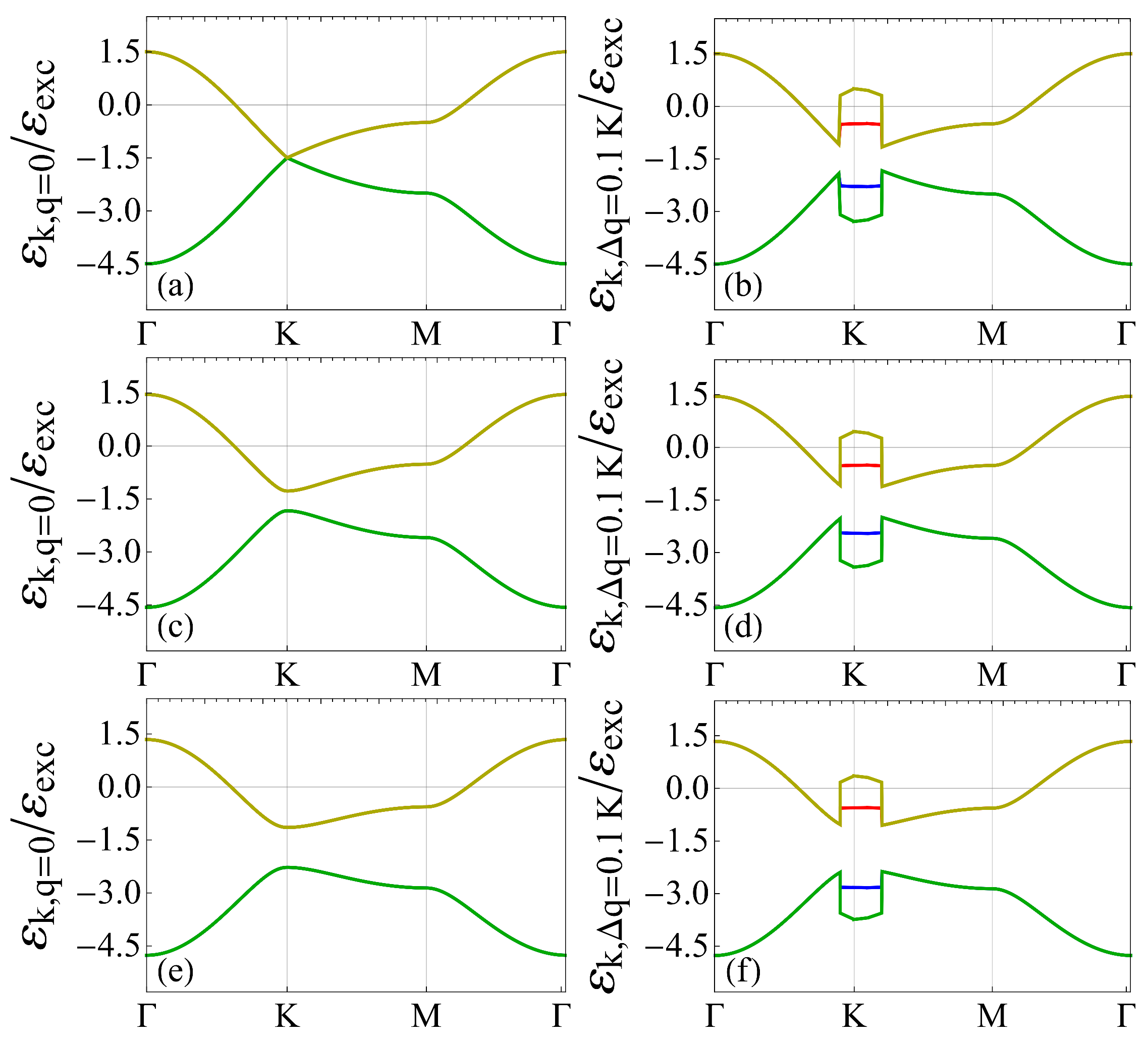}
\caption{\label{fig:Fig_12}(Color online) The electronic band structure in graphene at $V=0$ eV (see panels (a) and (b)), $V=0.5\gamma_0=1.5$ eV (see panels (c) and (d)) and $V=\gamma_0=3$ eV (see panels (e) and (f)). The phonon wave vector is set at the value ${\bf{q}}=0$ in the plots in panels (a), (c) and (e), while in the plots in panels (b), (d) and (f) it is changing with the electron wave vector ${\bf{k}}$ within the interval ${\bf{q}}\in\left[0.9{\bf{K}},{\bf{K}}\right]$. The local Hubbard potential has been fixed at $U=\gamma_0=3$ eV. The calculations have been performed in the strong electron-phonon coupling limit with $\lambda_{\rm eff}=0.21\gamma_0=0.63$ eV, and temperature has been set at the value $T=0.2\varepsilon_{\rm exc}=131.6$ K.} 
\end{figure} 
%

The band structure of graphene under consideration has been calculated in Fig.~\ref{fig:Fig_12}, for the case of half-filling with $\kappa=0.5$, corresponding to total average number of particles $n_{\rm fill}=2.0$. The local Hubbard-$U$ Coulomb potential has been fixed at $U=\gamma_0=3$ eV and strong electron-phonon coupling limit has been considered with $\lambda_{\rm eff}=0.21\gamma_0=0.63$ eV. The calculations have been done for the case $T=0.2\varepsilon_{\rm exc}=131.6$ K. The external gate-potential has been set at the values $V=0$ eV (see in panels (a) and (b)), $V=0.5\gamma_0=1.5$ eV (see panels (c) and (d)) and $V=\gamma_0=3$ eV (see panels (e) and (f)) and temperature was set at the value $T=0.2\varepsilon_{\rm exc}=131.6$ K. Moreover, two different limits of the phonon wave vector ${\bf{q}}$ have been considered. Mainly, the band structures in panels (a), (c) and (e) have been calculated for ${\bf{q}}=0$ and the band structures in panels (b), (d) and (f) for the case when ${\bf{q}}$ is varying in the interval ${\bf{q}}\in\left[0.9{\bf{K}},{\bf{K}}\right]$ (with $\Delta{\bf{q}}=0.1{\bf{K}}$). As we see, in Fig.~\ref{fig:Fig_12}, at zero phonon excitation, i.e., when ${\bf{q}}=0$, the calculations give the usual band structure of graphene \cite{cite_50} (see in panel (a)). At Dirac's point (see panel (a)), Fermi crossing energy is of order $\varepsilon_{\rm D}=-1.498\varepsilon_{\rm exc}=-84.9$ meV (see in panel (a)). At $V\neq 0$, the gate induced energy gaps were found of order $\varepsilon_{\rm g1}=31.7$ meV at $V=V_{\rm g1}=0.5\gamma_0=1.5$ eV (see in panel (c)), and $\varepsilon_{\rm g2}=64.1$ meV at $V=V_{\rm g2}=\gamma_0=3$ eV. Thus, practically, we have the relation 
\begin{eqnarray}
V_{\rm g2}/V_{\rm g1}=\varepsilon_{\rm g2}/\varepsilon_{\rm g1}.
\label{Equation_61}
\end{eqnarray}
We observe, in panel (b), in Fig.~\ref{fig:Fig_12}, that, in the case of unbiased graphene with $V=0$ eV, a very large superconducting band-gap $\varepsilon^{\rm sc}_{1\rm g}$ is opening at Dirac's point ${\bf{K}}$, between the quasiparticle energy branches $\varepsilon_{3{\bf{k}}}$ and $\varepsilon_{4{\bf{k}}}$. Namely, $\varepsilon^{\rm sc}_{\rm 1g}=\varepsilon_{2{\bf{k}}={\bf{K}}}-\varepsilon_{1{\bf{k}}={\bf{K}}}=1.78\varepsilon_{\rm exc}=101.3$ meV. In the picture, in panel (b), in Fig.~\ref{fig:Fig_12}, there is also two other phonon-mediated flat bands that are arising at the vicinity of Dirac point, within the interval $\Delta{\bf{k}}=\Delta{\bf{q}}=0.1{\bf{K}}$, indicating about the superconducting nature of those bands. The energy gap between these bands is of order $\varepsilon^{\rm sc}_{\rm 2g}=3.78\varepsilon_{\rm exc}=\varepsilon_{4{\bf{k}}={\bf{K}}}-\varepsilon_{3{\bf{k}}={\bf{K}}}=214.6$ meV. For the biased case, when $V=0.5\gamma_0$ (see in panel (d)), we get $\varepsilon^{\rm sc}_{\rm 1g}=1.93\varepsilon_{\rm exc}=\varepsilon_{2{\bf{k}}={\bf{K}}}-\varepsilon_{1{\bf{k}}={\bf{K}}}=109.8$ meV and $\varepsilon^{\rm sc}_{\rm 2g}=3.86\varepsilon_{\rm exc}=\varepsilon_{4{\bf{k}}={\bf{K}}}-\varepsilon_{3{\bf{k}}={\bf{K}}}=219.2$ meV. Furthermore, for higher value of $V$, namely, for the case $V=\gamma_0$ (see in panel (f), in Fig.~\ref{fig:Fig_12}), we have $\varepsilon^{\rm sc}_{\rm 1g}=2.27\varepsilon_{\rm exc}=\varepsilon_{2{\bf{k}}={\bf{K}}}-\varepsilon_{1{\bf{k}}={\bf{K}}}=128.7$ meV and $\varepsilon^{\rm sc}_{\rm 2g}=4.09\varepsilon_{\rm exc}=\varepsilon_{4{\bf{k}}={\bf{K}}}-\varepsilon_{3{\bf{k}}={\bf{K}}}=231.9$ meV. It is not difficult to realize that, for the unbiased case, when $V=0$ eV, the half of the difference between the band-gaps $\varepsilon^{\rm sc}_{\rm 2g}$ and $\varepsilon^{\rm sc}_{\rm 1g}$ is of order of single-particle ${\bf{q}}$-excitation energy $\varepsilon_{\rm exc}$, i.e.,
\begin{eqnarray}
\varepsilon_{\rm exc}=\frac{\varepsilon^{\rm sc}_{\rm 2g}-\varepsilon^{\rm sc}_{\rm 1g}}{2}.
\label{Equation_62}
\end{eqnarray}
In the matter of fact, the flat bands offer the possibility for the high temperature superconducting condensate state in the system \cite{cite_51, cite_52, cite_53, cite_54}. The obtained flat bands, in panels (b), (d) and (f), in Fig.~\ref{fig:Fig_12}, which originate from the phonons, promote the formation of the superconducting phase in our graphene system, the subject of which was the present paper.   

%
\begin{widetext}

\begin{figure}
	\includegraphics[scale=0.25]{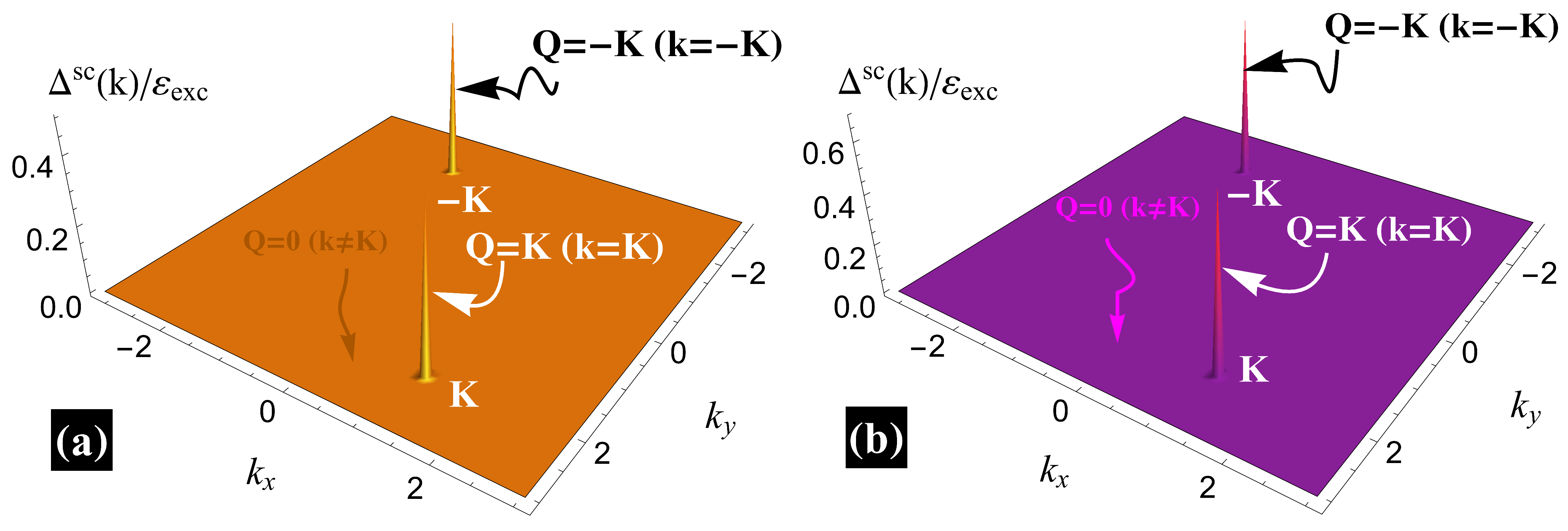}
	\caption{\label{fig:Fig_13}(Color online) The ${\bf{k}}$-map of the spin-triplet superconducting order parameter $\Delta^{\rm sc}\left({\bf{k}}\right)$, for two different temperatures: $T=0.12\varepsilon_{\rm exc}=78.95$ K (see in panel (a)) and $T=0.2\varepsilon_{\rm exc}=131.6$ K (see in panel (b)). The external gate-potential is set at the value $V=0$, and the local Hubbard-$U$ potential was fixed at $U=\gamma_0=3$ eV. The partial-filling case was considered with $\kappa=1.0$. The electron-phonon interaction energy is set at the value $\lambda_{\rm eff}=0.21\gamma_0=0.63$ eV.} 
\end{figure} 

\end{widetext}
%
%

Next, in Fig.~\ref{fig:Fig_13}, we have shown the ${\bf{k}}$-map for the triplet superconducting order parameter $\Delta^{\rm sc}$, at two different temperatures. We have fixed $T=0.12\varepsilon_{\rm exc}=78.95$ K (see in panel (a)) and $T=0.2\varepsilon_{\rm exc}=131.6$ K. The unbiased graphene, with $V=0$, has been considered. The local Hubbard-$U$ potential and the electron-phonon coupling parameter $\lambda_{\rm eff}$ have been set at the values $U=\gamma_0=3$ eV and $\lambda_{\rm eff}=0.21\gamma_0=0.63$ eV. 

Indeed, we supposed that the excitations with the wave vector ${\bf{q}}$ are present for the ${\bf{k}}$ states in the close vicinity of Fermi surface, where the chemical potential equals Fermi crossing energy of the bands $\mu=\varepsilon_{\rm D}$ (see in panel (a), in Fig.~\ref{fig:Fig_12}). This shift is indeed the result of the considered complex of interactions in the system which is somewhat similar with the discussion in \cite{cite_11}, where the shift of Fermi level and superconductivity in graphene have been attributed to the doping of graphene by metallic atoms. For the other states with ${\bf{k}}\neq {\bf{K}}$ the excitation vector is zero ${\bf{q}}=0$. In other words, we supposed that the external light source excitation, with the incident wave vector ${\bf{Q}}={\bf{q}}$ (see in Fig.~\ref{fig:Fig_1}), acts on the states in the close vicinity of the point ${\bf{K}}$ with $\Delta{{\bf{k}}}=0.1\left|{\bf{K}}\right|$, which has been indicated on the ordinate axes in the calculated band structures, in Fig.~\ref{fig:Fig_12}. We see, in Fig.~\ref{fig:Fig_13}, the presence of superconducting condensates peaks in the close vicinity of Dirac points ${\bf{k}}={\bf{K}}$ and ${\bf{k}}=-{\bf{K}}$. For the other electronic states, with ${\bf{k}}\neq{\bf{K}}$, or ${\bf{k}}\neq-{\bf{K}}$, the condensate peaks are absent. Moreover, the magnitudes of the condensate peaks get increased when augmenting temperature (see the result in panel (b), in Fig.~\ref{fig:Fig_13}), which is in complete agreement with the results for the superconducting order parameter $\Delta^{\rm sc}$ in Fig.~\ref{fig:Fig_5} (see the temperature dependence of plots, in Fig.~\ref{fig:Fig_5}). 
\section{\label{sec:Section_5} Concluding remarks}
%
We have considered spin-triplet superconductivity in gated graphene system, governed by the ${\bf{q}}$-excitation states. The effects of the external gate-potential and local Hubbard-$U$ interaction have considered. The superconducting state has been found at Dirac point by integrating out phonon field. In difference with the conventional root, a special scheme of decomposing the single-particle ${\bf{q}}$-excitation (above Fermi see) into two-particle $\bf{q}$-excitations (with calculated and predefined energies $\varepsilon_{\rm exc}=0.0567$ eV) has been proposed, and Nambu space vectors have been constructed, for this case. We obtained the system of self-consistent equations for the chemical potential, average charge density imbalance between $a$ and $b$ lattice sites and triplet superconducting order parameter. Then, we solved it exactly by employing Newton's fast convergence algorithm. We have found an unusual behavior of the triplet superconducting order parameter as a function of temperature and effective eletron-phonon interaction parameter. We have shown that the critical transition temperature $T_{\rm C1}$ for the triplet superconducting phase transition is surprisingly high, of order of $T_{\rm C1}=122$ K for the case of half-filling and in the case of the strong electron-phonon coupling energy. The second transition temperature was found of order of $T_{\rm C2}=236.24$ K, at which the system passes to the normal state. Those transition temperatures got decreased when passing to the partial-filling case (when the sum of average number of particles at different sublattice sites is fractional). In one of the partial filling cases, we obtain $T_{\rm C1}=60.3$ K and $T_{\rm c2}=148.6$ K, and the superconducting phase transition region gets narrowed in this case. Accordingly, the critical values have been found, for the effective electron-phonon interaction parameter, where the spin-triplet superconducting phase appear at the given temperature. Moreover, we have shown the dependence of the calculated parameters, namely the chemical potential, the average charge density imbalance and the superconducting order parameter, as a function of the external gate-potential and for different values of the local Hubbard-$U$ interaction parameter. We demonstrate that the in the large-$U$ limit, when the system is in the strong electron localization state, the superconducting phase gets enlarged, and the destruction of the superconducting order occurs at the higher critical values $V_{\rm C}$ of the external gate-potential. We have calculated the chemical potential and the average charge density difference for, both, superconducting and normal phase regions and we discussed the results. It is important to remark that the average charge density fluctuations are smaller for the large values of $U$ (Mott-Hubbard localization limit), which proof the localizing role of the Hubbard-$U$ interaction. Furthermore, the band structure of graphene has been calculated for different values of the applied gate-potential and ${\bf{q}}$-excitation vector. We have estimated the atomic relaxation time, during which the electrons at the vicinity of Fermi level with $\Delta{\bf{k}}=0.1{\bf{K}}$ acquire the phononic modes ${\bf{q}}_{i}={\bf{Q}}_{i}$, where $\bf{Q}_{i}$ is the wave number of the external radiation field. Moreover, the phonons, during that time, form a standing wave with the wave vector exactly coinciding with $\bf{Q}_{i}$. In the case when ${\bf{q}}_{i}=0$, we got the usual band structure of graphene, meanwhile, for the interval of wave number values $\Delta{\bf{q}}=\Delta{\bf{Q}}=0.1{\bf{K}}$, we obtained two different large band-gaps, even at $V=0$. Namely, in that region of the phononic excitations ${\bf{q}}_{i}$, the first band-gap $\varepsilon^{\rm sc}_{\rm 1g}$ was found between outermost quasi-flat energy bands and we get $\varepsilon^{\rm sc}_{\rm 1g}=101.3$ meV. The second band-gap $\varepsilon^{\rm sc}_{\rm 2g}$ was found between another pair of flat energy bands and we found it of order of $\varepsilon^{\rm sc}_{\rm 2g}=214.6$ meV. Those band-gap get slightly larger when increasing the gate potential $V$. Moreover, the difference of indicated two band-gaps, in the unbiased case, gives exactly the single-particle ${\bf{q}}$-excitation energy in the system $\varepsilon_{\rm exc}$, which reflects the fact that the triplet ${\bf{q}}$-superconductivity is mediated via two discrete single-particle ${\bf{q}}$-excitations, hypothesized in our model.

The results in the paper could be important for technological applications of monolayer graphene as a room-temperature spin-triplet superconducting material. 

\section*{Acknowledgments}

VA acknowledges MS for her useful comments on the subject of normalization of energy scales in the model and for her useful comments and suggestions on the calculations of the electronic band structure.   

\section*{Declarations}

\begin{itemize}
\item Conflict of interest/Competing interests (check journal-specific guidelines for which heading to use)

There is no conflict of interest between authors of present manuscript.

\item Availability of data and materials
 
This manuscript has no associated data or the data will not be deposited. [Authors' comment: This is a theoretical study and no experimental data.] 

\item Authors' contributions

VA developed the title of the manuscript and carried out theoretical and numerical calculations. MS participated in analyzing the obtained results and giving insights into some important principal subjects.

\end{itemize}
\appendix 
%
\section{\label{sec:Section_6} Calculation of coefficients in Eq.(\ref{Equation_55})}
%
In this Section, we give a detailed derivation of the coefficients $\alpha_{i{\bf{k}}}$, $\beta_{i{\bf{k}}}$ and $\gamma_{i{\bf{k}}}$ in Eq.(\ref{Equation_55}). For this purpose, as a precursing for further evaluations, we write each equation in the system of self-consistent equations, in Eq.(\ref{Equation_55}), in its proper definition-form
\begin{eqnarray}
&&\left\langle n_{\rm a}\left({\bf{r}}\tau\right)\right\rangle+\left\langle n_{\rm b}\left({\bf{r}}\tau\right)\right\rangle=\frac{1}{\kappa},
\nonumber\\
&&\left\langle n_{\rm a}\left({\bf{r}}\tau\right)\right\rangle-\left\langle n_{\rm b}\left({\bf{r}}\tau\right)\right\rangle=\delta{\bar{n}},
\nonumber\\
&&{\Delta}^{\rm sc}_{\eta,\sigma}=2g\lim_{\tau'\rightarrow \tau}\left\langle {\eta}_{\sigma}\left({\bf{r}}\tau'\right){\eta}_{\sigma}\left({\bf{r}}\tau\right)\right\rangle.
\label{Equation_A_1}
\end{eqnarray} 
In Fourier space, we have following definitions
\begin{eqnarray}
&&\left\langle n_{\rm a}\left({\bf{r}}\tau\right)\right\rangle=\frac{1}{2\beta^{2}{N}^{2}N_{\rm ph}}\sum_{\substack{{\bf{k}}{\bf{q}}\\ \nu_{n}\sigma}}\left\langle\bar{a}_{{\bf{k}}+\frac{\bf{q}}{2},\sigma}\left(\nu_{n}\right){a}_{{\bf{k}}+\frac{\bf{q}}{2},\sigma}\left(\nu_{n}\right)\right\rangle
\nonumber\\
&&+\left\langle\bar{a}_{{\bf{k}}-\frac{\bf{q}}{2},\sigma}\left(\nu_{n}\right){a}_{{\bf{k}}-\frac{\bf{q}}{2},\sigma}\left(\nu_{n}\right)\right\rangle,
\nonumber\\
&&\left\langle n_{\rm b}\left({\bf{r}}\tau\right)\right\rangle=\frac{1}{2\beta^{2}{N}^{2}N_{\rm ph}}\sum_{\substack{{\bf{k}}{\bf{q}}\\ \nu_{n}\sigma}}\left\langle\bar{b}_{{\bf{k}}+\frac{\bf{q}}{2},\sigma}\left(\nu_{n}\right){b}_{{\bf{k}}+\frac{\bf{q}}{2},\sigma}\left(\nu_{n}\right)\right\rangle
\nonumber\\
&&+\left\langle\bar{b}_{{\bf{k}}-\frac{\bf{q}}{2},\sigma}\left(\nu_{n}\right){b}_{{\bf{k}}-\frac{\bf{q}}{2},\sigma}\left(\nu_{n}\right)\right\rangle,
\nonumber\\
&&\Delta^{\rm sc}_{\rm \sigma}=\frac{2g}{\beta^{2}{N}^{2}N_{\rm ph}}\sum_{\substack{\eta=a,b\\{\bf{k}}{\bf{q}},\nu_{n}}}\left\langle \eta_{{\bf{k}}-\frac{{\bf{q}}}{2},\sigma}\left(\nu_{n}\right)\eta_{{\bf{k}}+\frac{{\bf{q}}}{2},\sigma}\left(\nu_{n}\right)\right\rangle.
\nonumber\\
\label{Equation_A_2}
\end{eqnarray}
Here, we will write the partition function of the system with generating source terms $J_{{\bf{k}}{\bf{q}},\sigma}\left(\nu_{n}\right)$ and $\bar{J}_{{\bf{k}}{\bf{q}},\sigma}\left(\nu_{n}\right)$, in order to calculate the averages in Eq.(\ref{Equation_A_2}). We have
\begin{eqnarray}
{J}_{{\bf{k}}{\bf{q}},\sigma}\left(\nu_{n}\right)&=\left(
\begin{array}{cccc}
\bar{J}_{a{\bf{k}}+\frac{{\bf{q}}}{2},\sigma}\left(\nu_{n}\right)\\ \\
{J}_{a{\bf{k}}-\frac{{\bf{q}}}{2},\sigma}\left(\nu_{n}\right)\\ \\
\bar{J}_{b{\bf{k}}+\frac{{\bf{q}}}{2},\sigma}\left(\nu_{n}\right)\\ \\
{J}_{b{\bf{k}}-\frac{{\bf{q}}}{2},\sigma}\left(\nu_{n}\right)
\end{array}\right),
\label{Equation_A_3}
\end{eqnarray}
and the partition function will be written as
\begin{widetext} 
\begin{eqnarray}
{\cal{Z}}&=&\int\left[{\cal{D}}\bar{\Psi}{\cal{D}}\Psi\right]e^{-\frac{1}{2\beta{N}N_{\rm ph}}\sum_{\substack{{\bf{k}}{\bf{q}} \\ \nu_{n}\sigma}}\bar{\Psi}_{{\bf{k}}{\bf{q}},\sigma}\left(\nu_{n}\right){\cal{G}}^{-1}_{{\bf{k}}{\bf{q}},\sigma}\left(\nu_{n}\right){\Psi}_{{\bf{k}}{\bf{q}},\sigma}\left(\nu_{n}\right)+\frac{1}{2\beta{N}N_{\rm ph}}\sum_{\substack{{\bf{k}}{\bf{q}} \\ \nu_{n}\sigma}}\bar{\Psi}_{{\bf{k}}{\bf{q}},\sigma}\left(\nu_{n}\right){J}_{{\bf{k}}{\bf{q}},\sigma}\left(\nu_{n}\right)+\frac{1}{2\beta{N}N_{\rm ph}}\sum_{\substack{{\bf{k}}{\bf{q}} \\ \nu_{n}\sigma}}\bar{J}_{{\bf{k}}{\bf{q}},\sigma}\left(\nu_{n}\right){\Psi}_{{\bf{k}}{\bf{q}},\sigma}\left(\nu_{n}\right)}
\nonumber\\
&&\approx e^{\frac{1}{2}\sum_{\substack{{\bf{k}}{\bf{q}}\\ \nu_{n}\sigma}}\bar{J}_{{\bf{k}}{\bf{q}},\sigma}\left(\nu_{n}\right){\cal{D}}_{{\bf{k}}{\bf{q}},\sigma}\left(\nu_{n}\right){J}_{{\bf{k}}{\bf{q}},\sigma}\left(\nu_{n}\right)}.
\label{Equation_A_4}
\end{eqnarray}
\end{widetext} 
Here, the ${\cal{D}}_{{\bf{k}}{\bf{q}},\sigma}\left(\nu_{n}\right)$ is the inverse of the matrix in Eq.(\ref{Equation_49}).   
The partition function in Eq.(\ref{Equation_A_4}) is, indeed, could be regarded as a generating functional of the complex source variables $J_{{\bf{k}}{\bf{q}},\sigma}\left(\nu_{n}\right)$ and $\bar{J}_{{\bf{k}}{\bf{q}},\sigma}\left(\nu_{n}\right)$
\begin{eqnarray}
{\cal{Z}}={\cal{Z}}\left[\bar{J}_{\sigma},J_{\sigma}\right].
\label{Equation_A_5}
\end{eqnarray}
After more, we perform the functional differentiation of both sides in the partition function in Eq.(\ref{Equation_A_4}), and we get 
\begin{eqnarray}
&&\frac{\delta^{2}{{\cal{Z}}}}{\delta{J}_{a{\bf{k}}+\frac{{\bf{q}}}{2},\sigma}({\bf{k}}\nu_{n})\delta{\bar{J}}_{a{\bf{k}}+\frac{{\bf{q}}}{2}},\sigma}
\nonumber\\
&&=-\frac{1}{4}\left\langle\bar{a}_{{\bf{k}}+\frac{{\bf{q}}}{2},\sigma}({\bf{k}}\nu_{n}){a}_{{\bf{k}}+\frac{{\bf{q}}}{2},\sigma}({\bf{k}}\nu_{n})\right\rangle=-\frac{1}{2}{\cal{D}}_{11{\bf{k}}{\bf{q}},\sigma}\left(\nu_{n}\right),	
\nonumber\\
\label{Equation_A_6}
\end{eqnarray}
\begin{eqnarray}
&&\frac{\delta^{2}{{\cal{Z}}}}{\delta{J}_{a{\bf{k}}-\frac{{\bf{q}}}{2},\sigma}({\bf{k}}\nu_{n})\delta{\bar{J}}_{a{\bf{k}}-\frac{{\bf{q}}}{2}},\sigma}
\nonumber\\
&&=-\frac{1}{4}\left\langle\bar{a}_{{\bf{k}}-\frac{{\bf{q}}}{2},\sigma}({\bf{k}}\nu_{n}){a}_{{\bf{k}}-\frac{{\bf{q}}}{2},\sigma}({\bf{k}}\nu_{n})\right\rangle=-\frac{1}{2}{\cal{D}}_{22{\bf{k}}{\bf{q}},\sigma}\left(\nu_{n}\right).	
\nonumber\\
\label{Equation_A_7}
\end{eqnarray}
Similarly, for the $b$-sublattice, we obtain
\begin{eqnarray}
&&\frac{\delta^{2}{{\cal{Z}}}}{\delta{J}_{b{\bf{k}}+\frac{{\bf{q}}}{2},\sigma}\left(\nu_{n}\right)\delta{\bar{J}}_{b{\bf{k}}+\frac{{\bf{q}}}{2},\sigma}\left(\nu_{n}\right)}
\nonumber\\
&&=-\frac{1}{4}\left\langle\bar{b}_{{\bf{k}}+\frac{{\bf{q}}}{2},\sigma}\left(\nu_{n}\right){b}_{{\bf{k}}+\frac{{\bf{q}}}{2},\sigma}\left(\nu_{n}\right)\right\rangle=-\frac{1}{2}{\cal{D}}_{33{\bf{k}}{\bf{q}},\sigma}\left(\nu_{n}\right),
\nonumber\\
\label{Equation_A_8}
\end{eqnarray}
\begin{eqnarray}
&&\frac{\delta^{2}{{\cal{Z}}}}{\delta{J}_{b{\bf{k}}-\frac{{\bf{q}}}{2},\sigma}({\bf{k}}\nu_{n})\delta{\bar{J}}_{b{\bf{k}}-\frac{{\bf{q}}}{2}},\sigma}
\nonumber\\
&&=-\frac{1}{4}\left\langle\bar{b}_{{\bf{k}}-\frac{{\bf{q}}}{2},\sigma}({\bf{k}}\nu_{n}){b}_{{\bf{k}}-\frac{{\bf{q}}}{2},\sigma}({\bf{k}}\nu_{n})\right\rangle=-\frac{1}{2}{\cal{D}}_{44{\bf{k}}{\bf{q}},\sigma}\left(\nu_{n}\right)
\nonumber\\
\label{Equation_A_9}
\end{eqnarray}
and, for the superconducting order parameters,
\begin{eqnarray}
&&\frac{\delta^{2}{{\cal{Z}}}}{\delta{\bar{J}}_{a{\bf{k}}-\frac{{\bf{q}}}{2},\sigma}({\bf{k}}\nu_{n})\delta{\bar{J}}_{a{\bf{k}}+\frac{{\bf{q}}}{2}},\sigma}
\nonumber\\
&&=-\frac{1}{4}\left\langle{a}_{{\bf{k}}-\frac{{\bf{q}}}{2},\sigma}({\bf{k}}\nu_{n}){a}_{{\bf{k}}+\frac{{\bf{q}}}{2},\sigma}({\bf{k}}\nu_{n})\right\rangle=-\frac{1}{2}{\cal{D}}_{12{\bf{k}}{\bf{q}},\sigma}\left(\nu_{n}\right).
\nonumber\\	
\label{Equation_A_10}
\end{eqnarray}
and
\begin{eqnarray}
&&\frac{\delta^{2}{{\cal{Z}}}}{\delta{\bar{J}}_{b{\bf{k}}-\frac{{\bf{q}}}{2},\sigma}({\bf{k}}\nu_{n})\delta{\bar{J}}_{b{\bf{k}}+\frac{{\bf{q}}}{2}},\sigma}
\nonumber\\
&&=-\frac{1}{4}\left\langle{b}_{{\bf{k}}-\frac{{\bf{q}}}{2},\sigma}({\bf{k}}\nu_{n}){b}_{{\bf{k}}+\frac{{\bf{q}}}{2},\sigma}({\bf{k}}\nu_{n})\right\rangle=-\frac{1}{2}{\cal{D}}_{34{\bf{k}}{\bf{q}},\sigma}\left(\nu_{n}\right).
\nonumber\\	
\label{Equation_A_11}
\end{eqnarray}
Then, we write Eq.(\ref{Equation_A_2})
\begin{widetext}
\begin{eqnarray}
&&\left\langle n_{\rm a}\left({\bf{r}}\tau\right)\right\rangle+\left\langle n_{\rm b}\left({\bf{r}}\tau\right)\right\rangle=\frac{1}{\beta^{2}{N}^{2}N_{\rm ph}}\sum_{\substack{{\bf{k}}{\bf{q}}\\ \nu_{n}\sigma}}\left({\cal{D}}_{11{\bf{k}}{\bf{q}},\sigma}\left(\nu_{n}\right)-{\cal{D}}_{22{\bf{k}}{\bf{q}},\sigma}\left(\nu_{n}\right)+{\cal{D}}_{33{\bf{k}}{\bf{q}},\sigma}\left(\nu_{n}\right)-{\cal{D}}_{44{\bf{k}}{\bf{q}},\sigma}\left(\nu_{n}\right)\right),
\nonumber\\
&&\left\langle n_{\rm a}\left({\bf{r}}\tau\right)\right\rangle-\left\langle n_{\rm b}\left({\bf{r}}\tau\right)\right\rangle=\frac{1}{\beta^{2}{N}^{2}N_{\rm ph}}\sum_{\substack{{\bf{k}}{\bf{q}}\\ \nu_{n}\sigma}}\left({\cal{D}}_{11{\bf{k}}{\bf{q}},\sigma}\left(\nu_{n}\right)-{\cal{D}}_{22{\bf{k}}{\bf{q}},\sigma}\left(\nu_{n}\right)-{\cal{D}}_{33{\bf{k}}{\bf{q}},\sigma}\left(\nu_{n}\right)+{\cal{D}}_{44{\bf{k}}{\bf{q}},\sigma}\left(\nu_{n}\right)\right),
\nonumber\\
&&\Delta^{\rm sc}=\frac{8}{\beta^{2}{N}^{2}N_{\rm ph}}\sum_{\substack{{\bf{k}}{\bf{q}}\\ \nu_{n}\sigma}}g_{\bf{q}}\left({\cal{D}}_{12{\bf{k}}{\bf{q}},\sigma}\left(\nu_{n}\right)+{\cal{D}}_{34{\bf{k}}{\bf{q}},\sigma}\left(\nu_{n}\right)\right).
\label{Equation_A_12}
\end{eqnarray}
\end{widetext}
From the other hand, we have 
\begin{eqnarray}
	&&{\cal{D}}_{ij{\bf{k}}{\bf{q}},\sigma}\left(\nu_{n}\right)=\beta{N_{\rm ph}N}\frac{{\cal{A}}_{ij{\bf{k}}{\bf{q}},\sigma}\left(\nu_{n}\right)}{\det{{\cal{D}}_{{\bf{k}}{\bf{q}}}\left(\nu_{n}\right)}}.
	\label{Equation_A_13}
\end{eqnarray}
The functions ${\cal{A}}_{ij{\bf{k}}{\bf{q}},\sigma}\left(\nu_{n}\right)$, in the right-hand side, in Eq.(\ref{Equation_A_13}), are the algebraic completions obtained after inverting the matrix in Eq.(\ref{Equation_49}). We have
\begin{widetext}
\begin{eqnarray}
&&{\cal{A}}_{11{\bf{k}}{\bf{q}},\sigma}\left(\nu_{n}\right)-{\cal{A}}_{22{\bf{k}}{\bf{q}},\sigma}\left(\nu_{n}\right)+{\cal{A}}_{33{\bf{k}}{\bf{q}},\sigma}\left(\nu_{n}\right)-{\cal{A}}_{44{\bf{k}}{\bf{q}},\sigma}\left(\nu_{n}\right)
\nonumber\\
&&={\cal{P}}^{\left(\mu\right)}_{{\bf{k}}{\bf{q}},\sigma}\left(\nu_{n}\right)=4\left(-i\nu_{n}\right)^{3}+\left(-i\nu_{n}\right)^{2}a_{1,\sigma}\left(\mu,U,V\right)+\left(-i\nu_{n}\right)b_{1{\bf{k}}{\bf{q}},\sigma}+c_{1{\bf{k}}{\bf{q}},\sigma},
\nonumber\\
&&{\cal{A}}_{11{\bf{k}}{\bf{q}},\sigma}\left(\nu_{n}\right)-{\cal{A}}_{22{\bf{k}}{\bf{q}},\sigma}\left(\nu_{n}\right)-{\cal{A}}_{33{\bf{k}}{\bf{q}},\sigma}\left(\nu_{n}\right)+{\cal{A}}_{44{\bf{k}}{\bf{q}},\sigma}\left(\nu_{n}\right)
\nonumber\\
&&={\cal{P}}^{\left(\delta{\bar{n}}\right)}_{{\bf{k}}{\bf{q}},\sigma}\left(\nu_{n}\right)=\left(-i\nu_{n}\right)^{2}a_{2,\sigma}\left(\mu,U,V\right)+\left(-i\nu_{n}\right)b_{2,\sigma}\left(\mu,U,V\right)+c_{2{\bf{k}}{\bf{q}},\sigma},
\nonumber\\
&&{\cal{A}}_{12{\bf{k}}{\bf{q}},\sigma}\left(\nu_{n}\right)={\cal{P}}^{\left(\Delta\right)}_{{\bf{k}}{\bf{q}},\sigma}\left(\nu_{n}\right)=-\Delta^{\rm sc}\left[\left(-i\nu_{n}\right)^{2}a_{3,\sigma}\left(\mu,U,V\right)+\left(-i\nu_{n}\right)b_{3,\sigma}\left(\mu,U,V\right)+c_{3{\bf{k}}{\bf{q}},\sigma}\right],
\label{Equation_A_14}
\end{eqnarray}
\end{widetext}
where
\begin{eqnarray}
&&a_{1,\sigma}\left(\mu,U,V\right)=-6\left(\mu_{\rm 1eff}+\mu_{\rm 2eff}\right),
\nonumber\\
&&b_{1{\bf{k}}{\bf{q}},\sigma}=2\left(2\left|\Delta^{\rm sc}\right|^{2}-\left|\gamma\left({\bf{k}}+\frac{q}{2}\right)\right|^{2}-\left|\gamma\left({\bf{k}}-\frac{q}{2}\right)\right|^{2}\right.
\nonumber\\
&&\left.+\mu^{2}_{\rm 1eff}+4\mu_{\rm 1eff}\mu_{\rm 2eff}+\mu^{2}_{\rm 2eff}\right),
\nonumber\\
&&c_{1{\bf{k}}{\bf{q}},\sigma}=-\left(\mu_{\rm 1eff}+\mu_{\rm 2eff}\right)\left[2\left|\Delta^{\rm sc}\right|^{2}-\left|\gamma\left({\bf{k}}+\frac{q}{2}\right)\right|^{2}\right.
\nonumber\\
&&\left.-\left|\gamma\left({\bf{k}}-\frac{q}{2}\right)\right|^{2}+2\mu_{\rm 1eff}\mu_{\rm 2eff}\right],
\nonumber\\
&&a_{2,\sigma}\left(\mu,U,V\right)=2\left(\mu_{\rm 1eff}-\mu_{\rm 2eff}\right),
\nonumber\\
&&b_{2,\sigma}\left(\mu,U,V\right)=-2\left(\mu^{2}_{\rm 1eff}-\mu^{2}_{\rm 2eff}\right),
\nonumber\\
&&c_{2,{\bf{k}}{\bf{q}},\sigma}=-\left(\mu_{\rm 1eff}-\mu_{\rm 2eff}\right)\left[2\left|\Delta^{\rm sc}\right|^{2}+\left|\gamma\left({\bf{k}}+\frac{q}{2}\right)\right|^{2}\right.
\nonumber\\
&&\left.+\left|\gamma\left({\bf{k}}-\frac{q}{2}\right)\right|^{2}-2\mu_{\rm 1eff}\mu_{\rm 2eff}\right],
\label{Equation_A_15}
\end{eqnarray}
and
\begin{eqnarray}
&&a_{3,\sigma}\left(\mu,U,V\right)\equiv 1, 
\nonumber\\
&&b_{3,\sigma}\left(\mu,U,V\right)=-2\mu_{\rm 2eff},
\nonumber\\
&&c_{3,{\bf{k}}{\bf{q}},\sigma}=\mu^{2}_{\rm 2eff}+\left|\Delta^{\rm sc}\right|^{2}+\gamma\left({\bf{k}}+\frac{q}{2}\right)\gamma\left({\bf{k}}-\frac{q}{2}\right).
\nonumber\\
\label{Equation_A_16}
\end{eqnarray}
In turn, $\det{{\cal{D}}_{{\bf{k}}{\bf{q}}}\left(\nu_{n}\right)}$, in Eq.(\ref{Equation_A_13}), is the determinant of the matrix ${\cal{D}}_{{\bf{k}}{\bf{q}},\sigma}\left(\nu_{n}\right)$ in Eq.(\ref{Equation_A_4}). It could be calculated as
\begin{eqnarray}
\det{{\cal{D}}_{{\bf{k}}{\bf{q}}}\left(\nu_{n}\right)}=\left(i\nu_{n}-\varepsilon_{1{\bf{k}}{\bf{q}}}\right)\left(i\nu_{n}-\varepsilon_{2{\bf{k}}{\bf{q}}}\right)\times
\nonumber\\
\times\left(i\nu_{n}-\varepsilon_{3{\bf{k}}{\bf{q}}}\right)\left(i\nu_{n}-\varepsilon_{4{\bf{k}}{\bf{q}}}\right),
\label{Equation_A_17}
\end{eqnarray}	
where the energies $\varepsilon_{i{\bf{k}}{\bf{q}}}$, with $i=1,...,4$, give the band-structure in the considered model. Furthermore, after putting back the expressions in Eq.(\ref{Equation_A_13}) into equations in Eq.(\ref{Equation_A_12}), and taking into account Eqs.(\ref{Equation_A_14}), (\ref{Equation_15}) and (\ref{Equation_A_16}), we perform the summations over the fermionic Matsubara frequencies. We obtain the system of self-consistent equations in Eq.(\ref{Equation_55}), where the coefficients $\alpha_{i{\bf{k}}}$, $\beta_{i{\bf{k}}}$ and $\gamma_{i{\bf{k}}}$ have the following expressions
\begin{eqnarray}
	\footnotesize
	\arraycolsep=0pt
	\medmuskip = 0mu
	\alpha_{i{\bf{k}}\sigma}
	=\left\{
	\begin{array}{cc}
		\displaystyle   \frac{\left(-1\right)^{i+1}}{\varepsilon_{1{\bf{k}}{\bf{q}}}-\varepsilon_{2{\bf{k}}{\bf{q}}}}\prod_{j=3,4}\frac{{\cal{P}}^{\left(\mu\right)}_{{\bf{k}}{\bf{q}},\sigma}\left(\nu_{n}\right)}{\left(\varepsilon_{i{\bf{k}}{\bf{q}}}-\varepsilon_{j{\bf{k}}{\bf{q}}}\right)}, \ \ \ $if$ \ \ \  i=1,2.
		\newline\\
		\newline\\
		{\footnotesize
			\arraycolsep=0pt
			\medmuskip = 0mu
			\begin{array}{cc}
				\displaystyle  & \frac{\left(-1\right)^{i+1}}{\varepsilon_{3{\bf{k}}{\bf{q}}}-\varepsilon_{4{\bf{k}}{\bf{q}}}}\prod_{j=3,4}\frac{{\cal{P}}^{\left(\mu\right)}_{{\bf{k}}{\bf{q}},\sigma}\left(\nu_{n}\right)}{\left(\varepsilon_{i{\bf{k}}{\bf{q}}}-\varepsilon_{j{\bf{k}}{\bf{q}}}\right)}, \ \ \ $if$ \ \ \  i=3,4,
		\end{array}}
	\end{array}\right.
\label{Equation_A_18}
\end{eqnarray}
\begin{eqnarray}
	\footnotesize
	\arraycolsep=0pt
	\medmuskip = 0mu
	\beta_{i{\bf{k}}\sigma}
	=\left\{
	\begin{array}{cc}
		\displaystyle   \frac{\left(-1\right)^{i+1}}{\varepsilon_{1{\bf{k}}{\bf{q}}}-\varepsilon_{2{\bf{k}}{\bf{q}}}}\prod_{j=3,4}\frac{{\cal{P}}^{\left(\delta{\bar{n}}\right)}_{{\bf{k}}{\bf{q}},\sigma}\left(\nu_{n}\right)}{\left(\varepsilon_{i{\bf{k}}{\bf{q}}}-\varepsilon_{j{\bf{k}}{\bf{q}}}\right)}, \ \ \ $if$ \ \ \  i=1,2.
		\newline\\
		\newline\\
		{\footnotesize
			\arraycolsep=0pt
			\medmuskip = 0mu
			\begin{array}{cc}
				\displaystyle  & \frac{\left(-1\right)^{i+1}}{\varepsilon_{3{\bf{k}}{\bf{q}}}-\varepsilon_{4{\bf{k}}{\bf{q}}}}\prod_{j=3,4}\frac{{\cal{P}}^{\left(\delta{\bar{n}}\right)}_{{\bf{k}}{\bf{q}},\sigma}\left(\nu_{n}\right)}{\left(\varepsilon_{i{\bf{k}}{\bf{q}}}-\varepsilon_{j{\bf{k}}{\bf{q}}}\right)}, \ \ \ $if$ \ \ \  i=3,4,
		\end{array}}
	\end{array}\right.
	\label{Equation_A_19}
\end{eqnarray}
\begin{eqnarray}
	\footnotesize
	\arraycolsep=0pt
	\medmuskip = 0mu
	\gamma_{i{\bf{k}}\sigma}
	=\left\{
	\begin{array}{cc}
		\displaystyle   \frac{\left(-1\right)^{i+1}}{\varepsilon_{1{\bf{k}}{\bf{q}}}-\varepsilon_{2{\bf{k}}{\bf{q}}}}\prod_{j=3,4}\frac{{\cal{P}}^{\left(\Delta\right)}_{{\bf{k}}{\bf{q}},\sigma}\left(\nu_{n}\right)}{\left(\varepsilon_{i{\bf{k}}{\bf{q}}}-\varepsilon_{j{\bf{k}}{\bf{q}}}\right)}, \ \ \ $if$ \ \ \  i=1,2.
		\newline\\
		\newline\\
		{\footnotesize
			\arraycolsep=0pt
			\medmuskip = 0mu
			\begin{array}{cc}
				\displaystyle  & \frac{\left(-1\right)^{i+1}}{\varepsilon_{3{\bf{k}}{\bf{q}}}-\varepsilon_{4{\bf{k}}{\bf{q}}}}\prod_{j=3,4}\frac{{\cal{P}}^{\left(\Delta\right)}_{{\bf{k}}{\bf{q}},\sigma}\left(\nu_{n}\right)}{\left(\varepsilon_{i{\bf{k}}{\bf{q}}}-\varepsilon_{j{\bf{k}}{\bf{q}}}\right)}, \ \ \ $if$ \ \ \  i=3,4.
		\end{array}}
	\end{array}\right.
	\label{Equation_A_20}
\end{eqnarray}
The polynomials ${\cal{P}}^{\left(\mu\right)}_{{\bf{k}}{\bf{q}},\sigma}\left(\nu_{n}\right)$, ${\cal{P}}^{\left(\delta{\bar{n}}\right)}_{{\bf{k}}{\bf{q}},\sigma}\left(\nu_{n}\right)$ and ${\cal{P}}^{\left(\Delta\right)}_{{\bf{k}}{\bf{q}},\sigma}\left(\nu_{n}\right)$, in Eqs.(\ref{Equation_A_18}), (\ref{Equation_A_19}) and (\ref{Equation_A_20}) have been defined in Eq.(\ref{Equation_A_14}), above. 
%
\section*{References}
%

\end{document}